\documentclass{article}
\usepackage{jheppub,esint,shuffle,psfrag}
 \usepackage[utf8]{inputenc}

\usepackage{varioref}
\usepackage{amsmath,amsfonts,amsthm,mathrsfs}
\usepackage{enumerate}
\usepackage{fancyvrb}
\usepackage{verbatim}
\usepackage{wrapfig}
\usepackage{appendix}
\usepackage{amstext}
\usepackage{amssymb}
\usepackage{graphicx}
\usepackage{color}
\usepackage{varioref}
\usepackage{multirow,graphics}
\usepackage{epstopdf}
\usepackage{bbm}
\usepackage{slashed}
\usepackage{psfrag}
\usepackage{relsize}
\usepackage{bbm}
\usepackage{mathrsfs}
\usepackage{epsfig}

\usepackage{mathtools}

       \usepackage{dutchcal}
       \usepackage{dsfont}

\DeclareMathAlphabet{\mathpzc}{OT1}{pzc}{m}{it}

\usepackage{tikz}
\usepackage{subcaption}

\usepackage{mathtools}
\usepackage{bigints}

\usetikzlibrary{decorations}

\usepackage{amstext}
\usepackage{amssymb}
\usepackage{amsmath}
\usepackage{verbatim}

\newcommand{\bunderline}[1]{\underline{#1\mkern-1mu}\mkern5mu }
\newcommand{\bunderlineB}[1]{\underline{#1\mkern-3mu}\mkern4mu }
\def\mr{{\bunderline{r}}}
\def\mR{{\bunderlineB{R}}}

\newcommand{\bea}{\begin{eqnarray}}
\newcommand{\eea}{\end{eqnarray}}
\def\overC{\overline{C}}

\def\e{ \textbf{e}  } 

\def\pa {\partial}
\def \nn {N}
\def\la{\label}

\def\tt{\top}
\def\nc{\mathcal{n}}

\def\rdr{ {\check{r}}  }
\def\qq{{\mathcal q}}
\def\gQ{{\mathpzc{Q}\hspace{.07em}}}
\def\mI{\mathbbm{1}}

\def\b {\beta}

\newcommand{\sfrac}[2]{{\textstyle\frac{#1}{#2}}}

\numberwithin{equation}{section}

\usepackage{tikz}
\usetikzlibrary{plotmarks,calc,decorations, decorations.pathmorphing, patterns}
\usetikzlibrary{arrows}
\tikzstyle dynkin node=[very thick,shape=circle,draw,inner sep=0pt,minimum size=5mm]
\tikzstyle dynkin line=[very thick]
\tikzstyle inverse line=[gray,line width=1.46pt,line cap=round, dash pattern=on 0pt off 2\pgflinewidth]
\tikzstyle red phase=[red,decoration={snake,amplitude=0.1mm,segment length=1.6mm},decorate]
\tikzstyle blue phase=[blue,decoration={snake,amplitude=0.1mm,segment length=0.9mm},decorate]
\tikzstyle green phase=[green,decoration={snake,amplitude=0.1mm,segment length=0.9mm},decorate]
\tikzstyle brown phase=[brown,decoration={snake,amplitude=0.1mm,segment length=0.9mm},decorate]
\newcommand{\boundellipse}[3]% center, xdim, ydim
{(#1) ellipse (#2 and #3)
}
\usetikzlibrary{decorations.pathmorphing}
\tikzstyle arrow=[thick,rounded corners=18pt,-latex]
\tikzstyle box=[draw,rounded corners,outer sep=4pt]
\tikzstyle B node=[outer sep=0pt]
\tikzstyle Q node=[inner sep=1pt,outer sep=0pt]
\definecolor{purple_nice}{rgb}{0.4,0.2,0.7}
\definecolor{fuel_blue}{RGB}{42,162,185}

\def\<{\langle}
\def\>{\rangle}

\makeatletter
\@ifundefined{usebibtex}{} {}
\makeatother

\def\Tr{\text{Tr}~}

\title{
\Large Quantum Trace Formulae for the Integrals of \\the Hyperbolic Ruijsenaars-Schneider model}

\author[a]{Gleb Arutyunov,}
\author[b]{~Rob Klabbers}
\author[a]{and\,~Enrico Olivucci}

\affiliation[a]{II. Institut f\"ur Theoretische Physik, Universit\"at Hamburg, Luruper Chaussee 149, 22761
Hamburg, Germany\\
Zentrum  f\"{u}r  Mathematische  Physik,  Universit\"{a}t  Hamburg,  Bundesstrasse  55,  20146  Hamburg,Germany}
\affiliation[b]{Institut f\"{u}r Physik und IRIS Adlershof, Humboldt-Universit\"{a}t zu Berlin, Zum Gro\ss en Windkanal 6, D-12489 Berlin, Germany}

\abstract{%Abstract:
We conjecture the quantum analogues of the classical trace formulae for the integrals of motion of the quantum hyperbolic Ruijsenaars-Schneider model.
This is done by departing from the classical construction where the corresponding model is obtained from the Heisenberg double by the Poisson reduction procedure.
We also discuss some algebraic structures associated to the Lax matrix in the classical and quantum theory which arise upon introduction of the spectral parameter.
}

\usepackage{esint}
\usepackage{breqn}
\def \Tr {\mathop{\rm Tr}\nolimits}

\def \e  {\mathop{\rm e}\nolimits}
\def \t {t}
\def \rr {\mathcal{r}}
\def \RR {\mathcal{R}}

\def\MM{{\mathcal M}}

\usepackage{upgreek}

\def\numberbysection{\@addtoreset{equation}{section}
                     \def\theequation{\thesection.\arabic{equation}}}

\begin{document}

\hskip9cm
\preprint{HU-EP-19/02,\;\;ZMP-HH/19-2}
\maketitle

\flushbottom

\newpage

\section{Introduction}
The Ruijsenaars-Schneider (RS) models \cite{Ruijsenaars:1986vq,Ruijsenaars:1986pp} continue to provide an outstanding theoretical laboratory for the study of various aspects 
of Liouville integrability, both at the classical and quantum level, see, for instance, \cite{Feher:2016nta, Feher:2018pmu, 2018arXiv180401766C, Chalykh:2018wce,Zabrodin:2014bpa, Grekov:2018htz}. Also, new interesting applications of these type of models were recently found in conformal field theories \cite{Isachenkov:2017qgn}.

%Also, new interesting applications of 
In this work we study some aspects related to the quantum integrability of the RS model with the hyperbolic potential.  
Recall that the definition of quantum integrability relies on the existence of a quantisation map
which maps a complete involutive family of classical integrals of motion into a set of commuting operators on a Hilbert space. 
In general, there are different ways to choose a functional basis for this involutive family which is mirrored by 
the ring structure of the corresponding commuting operators. In particular, a classical integrable structure, most 
conveniently encoded into a Lax pair $(L,M)$, produces a set of canonical integrals which are simply the eigenvalues 
of the Lax matrix. Their commutativity relies on the existence of the classical $r$-matrix \cite{Babelon:1990qk}.
Provided this matrix exists one can build up different classical involutive families represented, for instance,
by elementary symmetric functions of the eigenvalues of $L$  or, alternatively, by traces ${\rm Tr} L^k$ for $k\in {\mathbb Z}$.
Concerning the particular class of the RS hyperbolic models, the quantisation of a family of elementary symmetric functions associated to a properly chosen $L$ 
is well known and given by the Macdonald operators \cite{Ruijsenaars:1986pp, Mac}.
In this paper we conjecture the quantum analogues of ${\rm Tr} L^k$ built up in terms of the same $L$-operator 
that is used to generate  Macdonald operators through the determinant type formulae \cite{Has,Antonov:1997zc}. In fact there appear two commuting families $I_k^{\pm}$
that are given by the {\it quantum trace formulae}
\bea
\nonumber
I_k^{\pm} = \text{Tr}_{12}\left(C_{12}^{\t_2} L_1 \bar{R}_{21}^{\t_2} R_{\pm12}^{\t_2} L_1 \dots L_1 \bar{R}_{21}^{\t_2} R_{\pm12}^{\t_2} L_1 \right)\, ,
\eea
as quantisation of the classical integrals ${\rm Tr} L^k$.  In particular, $R$ and $\bar{R}$
are two quantum dynamical $R$-matrices that depend rationally on the variables $\gQ_i=e^{q_i}$, where $q_i$, $i=1,\ldots, N$ are coordinates, and satisfy a system of equations of Yang-Baxter type. 
Also, $R$ is a parametric solution of the standard quantum Yang-Baxter equation.\footnote{For the definition of other quantities, see the main text.} Departing from $I_k^{\pm}$ and introducing $\qq=e^{-\hbar}$, we then find that these integrals are related to the Macdonald operators ${\mathcal S}_k$ through the $\qq$-deformed analogues 
of the determinant formulae that in the classical case relate  the coefficients of characteristic polynomial of $L$ with invariants constructed out of ${\rm Tr}L^k$. The commutativity of \(I_k^{\pm}\) and their relation to Macdonald operators has been checked by explicit computation for sufficiently large values of \(N\).

We arrive to this expression for $I_k^{\pm}$ through the following chain of arguments. It is known that the Calogero-Moser-Sutherland models and their RS 
generalisations can be obtained at the classical level through the hamiltonian  or Poisson reduction applied to a system exhibiting free motion 
on one of the suitably chosen initial finite- or infinite-dimensional phase spaces \cite{KKS}-\cite{ACF}, \cite{Feher:2008moa,Feher:2009kp,Feher:2016nta,Feher:2018pmu}. For instance, the RS model with the 
rational potential is obtained by the hamiltonian reduction of the cotangent bundle $T^*G=G\ltimes \mathscr{G}$, where $G$ ia Lie group and $\mathscr{G}$ is its Lie algebra.  In  \cite{Arutyunov:1996cmb} the corresponding reduction was developed for the Lie group 
$G={\rm GL}(N,{\mathbb C})$ by employing a special parametrisation for the Lie algebra-valued element $\ell=TQT^{-1}\in \mathscr{G}$, where $Q$ is a diagonal matrix and $T$ is an element of the Frobenius group $F\subset G$. An analogous parametrisation is
used for the group element $g=UP^{-1}T^{-1}\in G$, where $U$ is Frobenius and $P$ is diagonal. If one writes $Q_i=q_i$ and $P_i=\exp p_i$, then $(p_i,q_i)$ is a system of canonical variables with the Poisson bracket 
$\{p_i,q_j\}=\delta_{ij}$.
In the new variables the Poisson structure of the cotangent bundle is then described in terms of the triangular dynamical matrix $r$ satisfying the classical Yang-Baxter equation (CYBE) and of another matrix $\bar{r}$. The cotangent bundle 
is easily quantised, in particular, the algebra of quantum $T$-generators is $T_1T_2=T_2T_1R_{12}$ and its consistency is guaranteed by the fact that the matrix $R$, being a quantisation of $r$, is triangular, $R_{12}R_{21}=\mI$,
and obeys the quantum Yang-Baxter equation. The quantum $L$-operator is then introduced as $L=T^{-1}gT$ and it is an invariant under the action of $F$. 
In \cite{Arutyunov:1996cmb} the same formula for $I_k$ as given above\footnote{In the rational case there is only one family, $R_{\pm 12}\to R_{12}$.} was derived by eliminating from the commuting operators ${\rm Tr}g^k={\rm Tr}TL^k T^{-1}$ an element $T$. 

To build up the hyperbolic RS model, one can start from the Heisenberg double  associated to a Lie group $G$.
As a manifold, the Heisenberg double is $G\times G$ and it has a well-defined Poisson structure 
being a deformation of the one on $T^*G$ \cite{SemenovTianShansky:1985my}. However, an attempt to repeat the same steps of the reduction procedure meets an obstacle: since 
the action of $G$ on the Heisenberg double is Poisson, rather than hamiltonian, the Poisson bracket of two Frobenius invariants, $\{L_1,L_2\}$, is not closed, {\it i.e.} it is not expressed via $L$'s alone. Moreover, for the same reason, the Poisson bracket \(\{p_i,p_j\}\) does not vanish on the Heisenberg double. On the other hand, a part of the non-abelian moment map 
generates second class constraints and to find the Poisson structure on the reduced manifold one has to resort to the Dirac bracket construction.\footnote{In \cite{Arutyunov:1996uw} this problem was avoided 
by looking at those entries of $L$ only that commute with the second class constraints.}  In this paper we work out the Dirac brackets for Frobenius invariants and show in detail how the cancellation of the non-invariant terms happens on the constraint surface. This leads to the canonical set of brackets for the degrees of freedom \((p_i,q_i)\) on the reduced manifold, the physical phase space of the RS model.
However, continuing along the same path as in the rational case \cite{Arutyunov:1996cmb} does not seem to yield $\{T,L\}$ and $\{T,T\}$ brackets. The variable $T$ is not invariant with respect to the stability subgroup of the moment map and computation of such brackets requires 
fixing a gauge, which makes the whole approach rather obscure. Moreover, the very simple and elegant bracket $\{L_1,L_2\}$ emerging on the reduced phase space looks the same as in the rational case, with one exception: now the $r$-matrix $r_{12}$ entering this bracket is not skew-symmetric, {\it i.e.} $r_{12}\neq r_{21}$.
We then  find a quantisation  of $r_{12}$: a simple quantum $R$-matrix $R_+$ satisfying $R_{+12}R_{-21}=\mI$, where $R_{-12}$ is another solution of the quantum Yang-Baxter equation.  In the absence of the triangular property for $R_{+12}$, assuming, for instance, the same algebra 
for $T$'s as in the rational case - that is $T_1T_2=T_2T_1R_{+12}$ -  would be inconsistent. Thus, at this point we simply conjecture that the integrals of the hyperbolic model have absolutely the same form as in the rational case, 
with the exception that the rational $R$-matrices are replaced by their hyperbolic analogues, which we explicitly find. That this conjecture yields integrals of motion can then be verified by tedious but direct computation and indeed holds true. Working out explicit expressions for these integrals for small numbers $N$ of particles we find the determinant formulae relating these integrals to the standard basis of Macdonald operators. The rest of the paper is devoted to the model whose formulation includes the spectral parameter. Neither for the rational nor for the hyperbolic case the spectral parameter 
is actually needed to demonstrate their Liouville integrability, but its introduction leads to interesting algebraic structures 
and clarifies the origin of the shifted Yang-Baxter equation \cite{Arutyunov:1996uw}
and its scale-violating solutions. 

The paper is organised as follows. In the next section we show how to obtain the hyperbolic RS model by the Poisson reduction of the Heisenberg double.
This includes the derivation of the Poisson algebra of the Lax matrix via the Dirac bracket construction. We also introduce the spectral parameter 
and build up the theory based on spectral parameter-dependent (baxterised) $r$-matrices. 
%We show that in spite of the fact that the constant $r$-matrices satisfy 
%the usual system of the CYBEs, their baxterised counterparts most naturally obey the CYBEs system but with extra terms involving the dilatation of 
%the spectral parameter.       
We also describe a freedom in the definition of $r$-matrices that does not change the Poisson algebra of $L$'s. In section  \ref{QMod}
we consider the corresponding quantum theory. Finding the hyperbolic quantum $R$-matrices $R_{\pm}$ and $\bar{R}$, we conjecture our main formula for the 
quantum integrals $I_k^{\pm}$ and explain how it is related to the basis of the Macdonald operators. The rest of the section is devoted to the 
quantum baxterised $R$-matrices and the quantum $L$-operator algebra. We show that in spite of the fact that the constant $R$-matrices satisfy the 
usual system of quantum Yang-Baxter equations, their baxterised counterparts instead obey its modification that involves rescalings
of the spectral parameter with the quantum deformation parameter $\qq=e^{-\hbar}$. Some technical details are relegated to two appendices. 
All considerations are done 
%for $G={\rm GL(N,{\mathbb C})}$ and they are
 in the context of holomorphic integrable systems. 

\section{The classical model from reduction}
\subsection{Moment map and Lax matrix}
We start with recalling the construction of the classical Heisenberg double associated to the group
$G={\text{GL} (N,{\mathbb C})}$. Let the entries of matrices $A,B\in G$ generate the coordinate ring of the algebra of functions on the Heisenberg double.
The Heisenberg double is a Poisson manifold with the following Poisson brackets 
\bea
\la{AB}
\begin{aligned}
&\{A_1,A_2\}=-\rr_-\, A_1A_2-A_1A_2\, \rr_+    +   A_1\, \rr_-A_2 + A_2\, \rr_+A_1\, , \\
&\{A_1,B_2\}=-\rr_-\, A_1B_2-A_1B_2\, \rr_-   +  A_1\, \rr_-B_2+B_2\, \rr_+A_1\, , \\
&\{B_1,A_2\}=-\rr_+\, B_1A_2-B_1A_2\, \rr_+   +  B_1\, \rr_-A_2+A_2\, \rr_+B_1\, , \\
&\{B_1,B_2\}=-\rr_-\, B_1B_2-B_1B_2\, \rr_+ + B_1\, \rr_-B_2  + B_2\, \rr_+B_1\, .
\end{aligned}
\eea
Here and elsewhere in the paper we use the standard notation where the indices $1$ and $2$ denote the different matrix spaces.
The matrix quantities $\rr_{\pm}$ are the following $r$-matrices 
\bea
\la{r-mat-gl}
\begin{aligned}
\rr_+&=+ \frac{1}{2}\sum_{i=1}^{\nn} E_{ii}\otimes E_{ii} +\sum_{i<j}^{\nn}E_{ij}\otimes E_{ji}\, , \\
\rr_-&=-\frac{1}{2}\sum_{i=1}^{\nn} E_{ii}\otimes E_{ii} -\sum_{i>j}^{\nn }E_{ij}\otimes E_{ji}\, ,
\end{aligned}
\eea
In the following we also need the split Casimir
\bea
 C=\sum_{i,j=1}^{N}E_{ij}\otimes E_{ji}\, ,
 \eea
whose action on the tensor product ${\mathbb C}^N\otimes {\mathbb C}^N$ is a permutation. In these formulae $E_{ij}$ stand for the standard matrix units.
The $r$-matrices (\ref{r-mat-gl}) satisfy the classical Yang-Baxter equation (CYBE) and have the following properties: $\rr_+ - \rr_-=C$ and $\rr_{\pm 21}=-\rr_{\mp 12}$.

The variables $(A,B)$ can be interpreted as a pair of monodromies of a flat connection on a punctured torus around its two fundamental cycles \cite{Fock:1998nu}.   
The monodromies are not gauge invariants as they undergo an adjoint action of the group of residual gauge transformations which coincides with $G$
\bea
\la{ad_double}
A\to hAh^{-1}\, , ~~~~B\to hBh^{-1}\, .
\eea
If $G$ is a Poisson-Lie group with the Sklyanin bracket 
\bea
\la{PL_G}
\{h_1,h_2\}=-[\rr_{\pm},h_1h_2]\, ,
\eea
then the transformations (\ref{ad_double}) are the Poisson maps for the structure (\ref{AB}). The non-abelian moment map $\MM$ of this action 
is given by 
\bea
\la{Mmon}
\MM=BA^{-1}B^{-1}A\, 
\eea
and it generates the following infinitesimal transformations of $(A,B)$
\bea
\la{MAB}
\begin{aligned}
&\{\MM_1,A_2\}=-(\rr_+\MM_1-\MM_1\rr_-) A_2 + A_2(\rr_+\MM_1-\MM_1\rr_-)\, , \\
&\{\MM_1,B_2\}=-(\rr_+\MM_1-\MM_1\rr_-) B_2 + B_2(\rr_+\MM_1-\MM_1\rr_-)\, .
\end{aligned}
\eea

To perform the reduction, we fix the moment map to the following value
\bea
\la{fixed_mom_trig}
\MM=\exp(\gamma\nc)\, ,
\eea 
where $\nc$ is the Lie algebra element
\bea
\la{v_m}
\nc=\e\otimes \e^{\t} - \mI \, ,    
\eea 
where $\e$ is an $N$-dimensional vector with all its entries equal to unity,  $ \e^\t=(1,\ldots, 1)$,
and $\gamma$ is a formal parameter which will be eventually interpreted as the coupling constant. 
Fixing this value of the moment map is posteriorly motivated by the fact that the dynamical model arising on the reduced space 
will have a close connection to the RS model we are after. 

We are thus led to find all $A,B$ that solve the following matrix equation  
\bea
\la{M_trigRS}
BA^{-1}B^{-1}A=e^{-\gamma}\mI - e^{-\gamma}\frac{1-e^{\nn\gamma}}{\nn} \e\otimes \e^\t \, ,
\eea
where on the right-hand side we worked out the explicit form of the exponential $\exp(\gamma\nc)$. In the following
we adopt the concise notation  
\bea
\la{constants_RS}
\omega=e^{-\gamma}\, , ~~~~\beta= - e^{-\gamma}\frac{1-e^{\nn \gamma}}{\nn }=-\frac{\omega}{\nn }(1-\omega^{-\nn})\, .
\eea
To solve (\ref{M_trigRS}), we introduce a convenient representation for $A$ and $B$:
\bea
\la{Trig_fact}
A&=&T\gQ T^{-1}\, , \\
\la{Trig_fact_1}
B&=&UP^{-1}T^{-1}\, .
\eea
Here $\gQ$ and $P$ are two diagonal matrices and $T,U\in G$ are two Frobenius matrices, {\it i.e.} they satisfy the Frobenius condition 
\bea
\la{Frob}
T\e=\e\, , ~~~~~~U\e=\e
\eea
and, therefore, belong to the Frobenius subgroup $F$ of $G$.

Introducing $W=T^{-1}U \in F$, equation  (\ref{M_trigRS}) takes the form
\bea
\la{moment_W}
\gQ^{-1}W^{-1}\gQ W= \omega \mI+\beta \e\otimes \e^\t U\, ,
\eea
where we used the fact that $U\in F$. Furthermore, we write
\bea
\nonumber
\gQ^{-1}W^{-1}\gQ - \omega W^{-1} =\beta \e\otimes \e^\t UW^{-1}=\beta \e\otimes \e^\t T\, .
\eea
This equation can be elementary solved for $W^{-1}$ and we get
%\bea
%W^{-1}_{ij} (\gQ_j/\gQ_i -\alpha)=\beta c_j
%\eea 
\bea
W^{-1}=\sum_{i,j=1}^{\nn}\frac{\beta }{\gQ_i^{-1}- \omega \gQ_j^{-1}}\frac{c_j}{\gQ_j} E_{ij}\, ,
\eea
where we introduced $c_j=(\e^\t T)_j$. The condition $W^{-1}\in F$ gives a set of equations to determine the coefficients $c_j$:
\bea
\nonumber
 \sum_{j=1}^{\nn} V_{ij}\frac{c_j}{\gQ_j} =1\, ,~~~~ \forall i\, .
\eea
Here $V$ is a Cauchy matrix with entries 
\bea
\nonumber
V_{ij}=\frac{\beta }{\gQ_i^{-1}- \omega \gQ_j^{-1}}\, .
\eea
We apply the inverse of $V$
\bea
\nonumber
V^{-1}_{ij}=\frac{1}{\beta(\gQ_i^{-1}-\omega^{-1} \gQ_j^{-1})} \frac{ \prod\limits_{a=1}^{\nn} (\omega \gQ_i^{-1}-\gQ_a^{-1})  }{ \prod\limits_{a\neq i}^{\nn}( \gQ_i^{-1}- \gQ_a^{-1})  } \frac{\prod\limits_{a=1}^{\nn} (\omega^{-1} \gQ_j^{-1}- \gQ_a^{-1}) }{  \prod\limits_{a\neq j}^{\nn} (\gQ_j^{-1}-\gQ_a^{-1})  } \, ,
\eea
to obtain the following formula for the coefficients $c_j$
\bea
\la{cj}
c_j=\gQ_j\sum_{j=1}^{\nn} V^{-1}_{ij}=\frac{(1-\omega)}{\beta }\, \frac{ \prod\limits_{a\neq j}^{\nn} ( \gQ_j^{-1}-\omega^{-1}\gQ_a^{-1})  }{ \prod\limits_{a\neq j}^{\nn} ( \gQ_j^{-1}- \gQ_a^{-1})  }
=\nn \frac{1-\omega}{1-\omega^{\nn}}\prod_{a\neq j}^{\nn}\frac{\gQ_j-\omega \gQ_a}{\gQ_j-\gQ_a}\, ,
%\frac{1}{\beta} \frac{ \prod\limits_{a=1}^n (\a \gQ_i^{-1}-\gQ_a^{-1})  }{ \prod\limits_{a\neq i}^n ( \gQ_i^{-1}- \gQ_a^{-1})  }\sum_j \frac{1}{(\gQ_i^{-1}-\a^{-1} \gQ_j^{-1})}\frac{\prod\limits_{a=1}^n (\a^{-1} \gQ_j^{-1}- \gQ_a^{-1}) }{  \prod\limits_{a\neq j}^n (\gQ_j^{-1}-\gQ_a^{-1})  } 
\eea
where we substituted $\beta$ from (\ref{constants_RS}).
Finally, inverting $W^{-1}$ we find $W$ itself
\bea
\la{W_trig}
W_{ij}(\gQ)=\frac{\gQ_i}{c_i}(V^{-1})_{ij}=\frac{ \prod\limits_{a\neq i}^{\nn} (\gQ_j^{-1}-\omega \gQ_a^{-1})  }{ \prod\limits_{a\neq j}^{\nn} ( \gQ_j^{-1}- \gQ_a^{-1})  }\, .
\eea
%In order to achieve the most similarity to the rational case, we introduce 
%\bea
%\la{b_trig}
%b_j(\gQ)=\sum_ic_i\gQ_iW_{ij}=\sum_i\gQ_i(V^{-1})_{ij}=\frac{\a}{\beta}\gQ_j\frac{ \prod\limits_{a=1}^n (\a^{-1} \gQ_j^{-1}-\gQ_a^{-1})  }{ \prod\limits_{a\neq j}^n ( \gQ_j^{-1}- \gQ_a^{-1})  }\, .
%\eea
It is obvious, that eq.(\ref{M_trigRS})  becomes equivalent to the following two constraints
\bea
\la{TU_trig}
U=TW(\gQ)\, , ~~~~ \e^\t T=c^\t\, ,
\eea
where $T,U\in F$, and the quantities $W(\gQ)$, $c(\gQ)$ are given by (\ref{W_trig}) and  (\ref{cj}), respectively. 
Any solution of $\e^\t T=c^\t$ can be constructed as $T=hT_0$, where $T_0$ is a particular solution of this equation and 
$h$ is a Frobenius group element which satisfies the additional constraint $\e^\t h=\e^\t$. In fact, the subgroup of $\boldsymbol{\digamma}\subset F\subset G$
determined by the conditions
\bea
  \boldsymbol{\digamma} = \{h\in G:~h\e=\e, ~~~\e^\t h=\e^\t \} \, ,
\eea  
constitutes the stability group\footnote{We do not include in $\boldsymbol{\digamma}$ the one dimensional dilatation subgroup  ${\mathbb C}^*\simeq \{h\in G:~h=c \mI, ~c\neq 0\}$, 
because its action on the phase space is not faithful.} of the moment map determined by the element $\nc$. Note that $\dim_{\mathbb{C}} F=N^2-N$ and $\dim_{\mathbb{C}}\boldsymbol{\digamma} =(N-1)^2$.

Now we can define a family of $G$-invariant dynamical systems\footnote{The systems whose hamiltonians are invariant under the action of $G$.} taking
the combination $L=W(\gQ)P^{-1}$ as their Lax matrix.
Explicitly, 
\bea
\la{Lax_red}
L=\sum_{i,j=1}^n\frac{(1-\omega)\gQ_i}{\gQ_i- \omega \gQ_j} \prod\limits_{a\neq j}^{\nn} 
\frac{\omega\gQ_j-\gQ_a}{\gQ_j-\gQ_a}P^{-1}_j E_{ij}\, .
\eea
After specifying the proper reality conditions, this $L$ becomes nothing else but the Lax matrix of the RS family with the hyperbolic potential.
Note that on the constrained surface the $A,B$-variables take the following form 
\bea
\nonumber
A(P,\gQ,h)=hT_0 Q T_0^{-1}h^{-1}\, , ~~~~B(P,\gQ,h)=hT_0 L T_0^{-1}h^{-1}\, , ~~~h\in \boldsymbol{\digamma}.
\eea 
The reduced phase space can be singled out by fixing the gauge to, for instance, $h=1$. Its dimension over $\mathbb{C}$ is $2N^2-(N^2-1)-\dim_{\mathbb{C}}\boldsymbol{\digamma} =2N$.

\subsection{Poisson structure on the reduced phase space}
Now we turn to the analysis of the Poisson structure of the reduced phase space. We 
find from (\ref{AB}) the following formula
\bea
\la{QB}
\{\gQ_j,B\}= B\sum_{kl}T_{lj}\gQ_jT^{-1}_{j k}E_{lk}\, .
\eea
Next, we need to determine the bracket between $\gQ_j$ and $P_i$. We have 
\bea
\nonumber
\{\gQ_j, P_i \}=\frac{\delta P_i}{\delta A_{ mn} } \{ \gQ_j, A_{mn} \}+\frac{\delta P_i}{\delta B_{ mn}}\{\gQ_j, B_{mn}\}\, .
\eea
Here the first bracket on the right-hand side vanishes because all $\gQ_j$ commute with $A$.\footnote{The spectral invariants of $A$ are central in the Poisson subalgebra of $A$, the latter is described by the Semenov-Tian-Shansky bracket \cite{SemenovTianShansky:1985my}
given by the first line in (\ref{AB}).}    
To compute the second bracket, we consider the variation of $B=UP^{-1}T^{-1}$
\bea
\nonumber
U^{-1}\delta B\,  TP=U^{-1}\delta U-P^{-1}\delta P\, . 
\eea
Note that this formula does not include the variation $\delta T$. This is because  $T$ is solely determined by $A$, so so is its variation. 
The condition $\delta U\e=0$ allows one to find 
\bea
\nonumber
\frac{\delta P_i}{\delta B_{mn}}=-\sum_r P_i U_{im}^{-1}(TP)_{nr}\, .
\eea
 We thus have
 \bea
 \la{QP}
 \{\gQ_j, P_i \}=-  \sum_r P_i U_{im}^{-1}(TP)_{nr}(BT)_{mj}\gQ_jT^{-1}_{jn}=- \gQ_i P_i \delta_{ij}\, ,
 \eea
and similarly one can check the bracket \(\{\gQ_i,\gQ_j\}=0\).
These formulae suggests to employ the exponential parametrisation for both $P$ and $\gQ$, that is, to set 
$$
P_i=\exp p_i\, , ~~~~~~\gQ_i=\exp q_i\, , 
$$ where $(p_i,q_i)$ satisfy the canonical relations $\{p_i,q_j\}=\delta_{ij}$.

An $\boldsymbol{\digamma} $-invariant extension of the Lax matrix away from the reduced phase space is naturally given by 
the following Frobenius invariant  
\bea
\la{Trig_Lax}
L=T^{-1}BT\, , 
\eea
where $T$ is an element of the Frobenius group entering the factorisation (\ref{Trig_fact}).
The Poisson bracket of $\gQ_j$ with components of $L$ is computed in a straightforward manner 
$$
\{\gQ_j,L_{mn}\}=\{\gQ_j,(T^{-1}BT)_{mn}\}=\sum_p(T^{-1}B)_{mp}\sum_{kl}T_{lj}\gQ_jT^{-1}_{j k}(E_{lk})_{ps}T_{sn}= L_{mn}\gQ_n\delta_{jn}\, ,
$$  
which is perfectly compatible with the form (\ref{Lax_red}) of the Lax matrix on the reduced space. In matrix form the previous formula reads as
\bea
\la{bracket_LQ}
\{\gQ_1,L_2\}=\gQ_1L_2\overC_{12}\, ,~~~~\bar{C}_{12}=\sum_{j=1}^N E_{jj}\otimes E_{jj}\, .
\eea

As to the brackets between the entries of $L$, this time they cannot be represented in terms of $L$ alone but also involve $T$.  
Ultimately, such a structure is a consequence of the fact that the action of the Poisson-Lie group $G$ on the phase space is Poisson rather than hamiltonian, so that there is an obstruction for the Poisson bracket of two Frobenius invariants to also be such an invariant. 
In addition, computing the Dirac brackets of $L$ one cannot neglect a non-trivial contribution from the second class constraints and, therefore, the analysis of the Poisson structure for $L$ 
requires, as an intermediate step, to understand the nature of the constraints (\ref{M_trigRS}) imposed in the process of reduction. The same argument holds for the Poisson brackets between any of the Frobenius invariants \(W=T^{-1}U\) and \(P\), showing as a particular case that $P_i$'s have a non-vanishing Poisson algebra on the Heisenberg double.\footnote{At the level of quantisation, this fact prevents one from obtaining the quantum RS model starting from the algebra of the quantum Heisenberg double. Indeed, doing so one should later restore the canonical commutation relations of \((P,Q)\) sub-algebra by imposing an analogue of the Dirac constraints at the quantum level.}
We save the details of the corresponding analysis for appendix  \ref{app:cauchy} and present  here
the final result for the Poisson bracket between the entries of the Lax matrix on the reduced phase space
\bea
\la{LL_trig}
\{L_1, L_2\}&=& r_{12} L_1 L_2- L_1 L_2\mr_{12}
%-\bar{r}_{12}(\gQ)+\bar{r}_{21}(\gQ)  )\\
%\nonumber
+ L_1\bar{r}_{21} L_2- L_2\bar{r}_{12} L_1\, .
\eea
Clearly, the bracket (\ref{LL_trig}) has the same form as the corresponding bracket  for the rational RS model \cite{Arutyunov:1996cmb} albeit with new dynamical $r$-matrices 
for which we got the following explicit expressions \footnote{The quadratic and linear forms of the \(r\)-matrix structure for the RS model have been investigated in \cite{Suris:1996mg,Avan:1995dk,Babelon:1993bx,Nijhoff:1996pr,Sklyanin:1993uj}.}
%\footnote{To simplify the notation, we often omit the argument $\gQ$ of the $r$-matrices.}
\bea
\la{r_trig}
\begin{aligned}
r &=
\sum_{i\neq j}^{\nn}\Big(\frac{\gQ_j}{\gQ_{ij}} E_{ii}-\frac{\gQ_i}{\gQ_{ij}} E_{ij}\Big)\otimes (E_{jj}-E_{ji})\, , \\
\bar{r} &=\sum_{i\neq j}^{\nn}\frac{\gQ_i}{\gQ_{ij}}(E_{ii} - E_{ij})\otimes  E_{jj}\, , \\
%\la{mr_trig}
\mr&=\sum_{i\neq j}^{\nn}\frac{\gQ_i}{\gQ_{ij}}(E_{ij}\otimes E_{ji}-E_{ii}\otimes E_{jj})\, ,
 \end{aligned}
\eea
where we introduced the notation $\gQ_{ij}=\gQ_i-\gQ_j$. 
This structure can be obtained as well after the computation of the Dirac brackets of \(W\) and \(P\) on the reduced phase space
\begin{align}
\label{WW}
\{W_1,W_2\}=&\,[r_{12},W_1 W_2]\\
\label{WP}
\{W_1,P_2\}=&\,[\bar r_{12},W_1] P_2\\
\label{PPred}
\{P_1,P_2\}=&\,0\, ,
\end{align}
using the decomposition \(L=WP^{-1}\). Remarkably the imposition of Dirac constraints makes the Poisson subalgebra \(\{P_i\}\) abelian, allowing the interpretation of components \(p_i=\log P_i\) as particle momenta.
Concerning the properties of the matrices \eqref{r_trig} and the Lax matrix, we note the following: first, $\mr$ is expressed via $r$ and $\bar{r}$ as 
\bea
\la{mr}
\mr_{12}=r_{12}+\bar{r}_{21}-\bar{r}_{12}\, .
\eea
Second, the matrix $r$ is degenerate, $\det\,  r=0$, and it obeys the characteristic equation $r^2=-r$. 
Moreover, in contrast to the rational case \cite{Arutyunov:1996cmb}, $r$ is not symmetric, rather it has the property  
\bea 
\la{prop_r}
r_{12}+r_{21}=C_{12}-\mI\otimes \mI\, .
\eea
%where $\mI_{12}$ is the identity matrix.
%Third, the $r$-matrices (\ref{r_trig}) obey the same equations (\ref{r_eq_1}), (\ref{r_eq_2}) and (\ref{YB_mr}). 
Third, it is a matter of 
straightforward calculation to verify that the Lax matrix (\ref{Lax_red}) obeys the Poisson algebra relations (\ref{LL_trig}), provided 
the bracket between the components of $\gQ$ and $P$ is given by (\ref{QP}),\eqref{PPred}.
Finally,  as a consequence of the Jacobi identities, the matrices (\ref{r_trig}) satisfy a system of equations of Yang-Baxter type.
In particular, for $r$ one has just the standard CYBE
\bea
\la{r_eq_1}
[r_{12},r_{13}]+[r_{12},r_{23}]+[r_{13},r_{23}]=0\, .
\eea
In addition, there are two more equations involving $r$ and $\bar{r}$
\bea
\la{r_eq_2}
\begin{aligned}
&[\bar{r}_{12},\bar{r}_{13}]+\{\bar{r}_{12},p_3\}-\{\bar{r}_{13},p_2\}=0\, ,\\
&[r_{12},\bar{r}_{13}]+[r_{12},\bar{r}_{23}]+[\bar{r}_{13},\bar{r}_{23}]+\{r_{12},p_3\}=0\, .
\end{aligned}
\eea  
The matrix $\mr$ satisfies the classical analogue of the Gervais-Neveu-Felder equation \cite{Gervais:1983ry,Felder:1994pb}
\bea
\la{YB_mr}
[\mr_{12},\mr_{13}]+[\mr_{12},\mr_{23}]+[\mr_{13},\mr_{23}]+\{\mr_{12},p_3\}-\{\mr_{13},p_2\}+\{\mr_{23},p_1\}=0\, .
\eea 

It is elementary to verify that the quantities 
\bea
\la{ctr}
I_k={\rm Tr}L^k
\eea 
are in involution with respect to (\ref{r_trig}). This property of $I_k$ is, of course, inherited from   
the same property for ${\rm Tr}B^k$ on the original phase space (\ref{AB}). We refer to (\ref{ctr}) as the \emph{classical trace formula}.
%The integrals $I_k$ arise as entries in the determinate formula 

\subsection{Introduction of a spectral parameter}
Here we introduce a Lax matrix depending on a spectral parameter and discuss the associated algebraic structures 
and an alternative way to exhibit commuting integrals. 

To start with, we point out one important identity satisfied by the Lax matrix (\ref{Lax_red}).
According to the moment map equation (\ref{moment_W}), we have
\bea
\la{Wpr}
\omega\gQ^{-1}W\gQ =W \Big[ \mI+\frac{\beta}{\omega}\, \e\otimes \e^\t U\Big]^{-1}\, ,
\eea
The inverse on the 
right-hand side of the last expression can be computed with the help of the well-known Sherman-Morrison formula
and we get 
\bea
\la{Wpr1}
\omega\gQ^{-1}W\gQ =W\Big[  \mI-\frac{1-\omega^N}{N} \e\otimes \e^\t U \Big]=W-\frac{1-\omega^N}{N} \e\otimes\, c^\t W\, ,
\eea
where we used the fact that $W$ is a Frobenius matrix, so that $W\e=\e$. 
Here the vector $c$ has components (\ref{cj}) and satisfies the relation $\e^\t T=c^\t$. Multiplying both sides
of (\ref{TU_trig}) with $P^{-1}$ we obtain the following identity
\bea
\la{Lax_ident}
\omega\gQ^{-1}L\gQ=L-\frac{1-\omega^N}{N} \e\otimes c^\t L\, ,
\eea 
for the Lax matrix (\ref{Lax_red}). 

Evidently,  we can consider 
\bea
\la{new_Lax}
L'=\omega\gQ^{-1}L\gQ
\eea
as another Lax matrix since the evolution equation of the latter is of the Lax form
\bea
\dot{L}'=[M'\, , L']\, , ~~~~M'=\gQ^{-1}M\gQ-\gQ^{-1}\dot{\gQ}\, ,
\eea 
where $M$ is defined by the hamiltonian flow of $L$. Note that one can add to $M'$ any function of $L'$ without changing the evolution equation for $L'$,
which defines a class of equivalent $M'$'s.  Now, it turns out that due to the special dependence of $L$ on the momentum, $M$ and $M'$ fall in the same 
equivalence class.  To demonstrate this point, it is enough to consider the simplest hamiltonian \(H  = \Tr L\)  for which the matrix
\(M\) is given by
\bea
\la{M_k=1}
M=\sum_{i\neq j}^{\nn} \frac{\gQ_j}{\gQ_{ij}}L_{ij}(E_{ii}-E_{ij})\, ,
\eea
It follows from (\ref{bracket_LQ}) that for the flow generated by this hamiltonian
\bea
\nonumber
\gQ^{-1}\dot{\gQ}=\gQ^{-1}\{H,\gQ\}=-\sum_{i=1}^{\nn}L_{ii}E_{ii}\, .
\eea
Therefore, 
\bea
\nonumber
M'=\gQ^{-1}M\gQ-\gQ^{-1}\dot{\gQ}=\sum_{i\neq j}^{\nn}\frac{\gQ_j}{\gQ_{ij}}L_{ij}\Big(E_{ii}-\frac{\gQ_j}{\gQ_i}E_{ij}\Big)+\sum_{i=1}^{\nn}L_{ii}E_{ii}\, .
\eea
 Taking into account that $\gQ_j/(\gQ_{ij}\gQ_i)=1/\gQ_{ij}-1/\gQ_i$, we then find
%\bea
%\nonumber
%\frac{\gQ_j}{\gQ_{ij}\gQ_i}=\frac{1}{\gQ_{ij}}-\frac{1}{\gQ_i}\, ,
%\eea
%we will get 
\bea
\nonumber
M'=\sum_{i\neq j}^{\nn}\frac{\gQ_j}{\gQ_{ij}}L_{ij}(E_{ii}-E_{ij})+\sum_{i\neq j}^{\nn}\gQ_i^{-1}L_{ij}\gQ_jE_{ij}
+\sum_{i=1}^{\nn}L_{ii}E_{ii}=M+L'\, .
\eea
Hence, $M'$ is in the same equivalence class as $M$ and, therefore, we can take the dynamical matrix $M$ to be the same for both $L$ and $L'$.  

The above observation motivates to introduce a Lax matrix depending on a spectral parameter just as a linear combination of $L$ and $L'$. Namely, we can define
\bea
\la{spec_dep_L}
L(\lambda)=L-\frac{1}{\lambda}L'\, ,
\eea
where $\lambda\in {\mathbb C}$ is the spectral parameter. The matrix $L(\lambda)$ has a pole at zero and the original matrix $L$ is obtained from $L(\lambda)$
in the limit $\lambda\to \infty$, in particular,
\bea
\la{HLLp}
H=\lim_{\lambda\to\infty} {\rm Tr }L(\lambda)={\rm Tr} L\, .
\eea
The evolution equation for $L(\lambda)$ must, therefore, be of the form 
\bea
\la{evolLLpr}
\dot{L}(\lambda)=\{H,L(\lambda)\}=[M , L(\lambda)]\, ,
\eea
where $M$ is the expression (\ref{M_k=1}).

The next task is to compute the Poisson brackets between the components of  (\ref{spec_dep_L}).  We aim at finding
a structure similar to (\ref{LL_trig}), namely,
\bea
\begin{aligned}
\label{LLsp}
\{L_1(\lambda),L_2(\mu)\}&=
 r_{12}(\lambda,\mu) L_1(\lambda) L_2(\mu)- L_1(\lambda)  L_2(\mu)\mr_{12}(\lambda,\mu) \\
\la{sdL_t}
 &\hspace{2cm}+ L_1(\lambda) \bar{r}_{21}(\mu)L_2(\mu)- L_2(\mu)\bar{r}_{12}(\lambda) L_1(\lambda) \, ,
 \end{aligned}
\eea
where $ r(\lambda,\mu)$, $ \mr(\lambda,\mu)$ and $\bar{r}(\lambda)$ are some spectral-parameter-dependent $r$-matrices. We show how to 
derive these $r$-matrices in appendix  \ref{app:cauchy}. Our considerations are essentially based on the identity (\ref{Lax_ident}). To state the corresponding result, we need the matrix
\bea
\la{sigma_mat}
\sigma_{12}=\sum_{i\neq j}^{\nn} (E_{ii}-E_{ij})\otimes E_{jj}\, .
\eea
The {\it minimal solution}\footnote{The explanation of its minimal character will be given later.} for the spectral-dependent $r$-matrices realising the Poisson algebra (\ref{sdL_t}) is then found to be
\bea
\la{spec_dep_r_final}
\begin{aligned}
&r_{12}(\lambda,\mu)=\frac{\lambda r_{12}+\mu r_{21}}{\lambda-\mu}+\frac{\sigma_{12}}{\lambda-1} -\frac{\sigma_{21}}{\mu-1} \, ,\\
&\bar{r}_{12}(\lambda)=\bar{r}_{12}+\frac{\sigma_{12}}{\lambda-1} \, , \\
&\mr_{12}(\lambda,\mu)=r_{12}(\lambda,\mu)+\bar{r}_{21}(\mu)-\bar{r}_{12}(\lambda)=\frac{\lambda \mr_{12}+\mu \mr_{21}}{\lambda-\mu}\, .
\end{aligned}
\eea
The matrices $r$ and $\mr$ are skew-symmetric in the sense that
\bea
r_{12}(\lambda,\mu)=-r_{21}(\mu,\lambda)\, , ~~~~\mr_{12}(\lambda,\mu)=-\mr_{21}(\mu,\lambda)\, .
\eea

Further, one can establish implications of the Jacobi identity satisfied by (\ref{sdL_t}) for these $r$-matrices.
Introducing the dilatation operator acting on the spectral parameter 
$$
D_{\lambda}= \lambda \frac{\pa}{\pa \lambda}\, ,
$$
we find that  the $r$-matrix $r(\lambda,\mu)$ does not satisfy the standard CYBE but rather the following modification thereof 
\bea
\la{shiftedCYB}
&&\hspace{-1cm}[r_{12}(\lambda,\mu),r_{13}(\lambda,\tau)]+[r_{12}(\lambda,\mu),r_{23}(\mu,\tau)]+[r_{13}(\lambda,\tau),r_{23}(\mu,\tau)]=\, \\
\nonumber
&&~~~~=-\,(D_{\lambda}+D_{\mu}) r_{12}(\lambda,\mu)+(D_{\lambda}+D_{\tau}) r_{13}(\lambda,\tau)
-(D_{\tau}+D_{\mu}) r_{23}(\mu,\tau)\, .
\eea
Following \cite{ACF}, we refer to (\ref{shiftedCYB}) as the {\it shifted classical Yang-Baxter equation}. This equation can be rewritten in the form of the standard Yang-Baxter equation 
\bea
\nonumber
[\hat{r}_{12}(\lambda,\mu),\hat{r}_{13}(\lambda,\tau)]+[\hat{r}_{12}(\lambda,\mu),\hat{r}_{23}(\mu,\tau)]+[\hat{r}_{13}(\lambda,\tau),\hat{r}_{23}(\mu,\tau)]=0\, .
\eea
for the matrix differential operator 
\bea
\hat{r}(\lambda,\mu)=r(\lambda,\mu)-D_{\lambda}+D_{\mu}\, .
\eea
 There are also two more equations involving the matrix $\bar{r}$
\bea
\la{shifted1}
&&\hspace{-1cm}[r_{12}(\lambda,\mu),\bar{r}_{13}(\lambda)+\bar{r}_{23}(\mu)]+[\bar{r}_{13}(\lambda),\bar{r}_{23}(\mu)]+P_3^{-1}\{r_{12}(\lambda,\mu),P_3\}=\\
\nonumber
&&\hspace{3cm}=-(D_{\lambda}+D_{\mu})r_{12}(\lambda,\mu)+(D_{\lambda}\bar{r}_{13}(\lambda)-D_{\mu}\bar{r}_{23}(\mu))\, 
\eea
and 
\bea
\la{shifted2}
[\bar{r}_{12}(\lambda),\bar{r}_{13}(\lambda)]+P_3^{-1}\{\bar{r}_{12}(\lambda),P_3\}-P_2^{-1}\{\bar{r}_{13}(\lambda),P_2\}=-D_{\lambda}(\bar{r}_{12}(\lambda)-\bar{r}_{13}(\lambda))\, .
\eea
One can check that relations (\ref{shiftedCYB}), (\ref{shifted1})  and (\ref{shifted2}) guarantee the fulfilment 
of the Jacobi identity for the brackets (\ref{bracket_LQ}) and (\ref{LL_trig}). Note that $\mr$ is scale-invariant: $(D_{\lambda}+D_{\mu})\mr(\lambda,\mu)=0$, implying that 
it depends on the ratio $\lambda/\mu$.  This property does not hold, however, for $r$ and $\bar{r}$.

The solution we found for the spectral-dependent dynamical $r$-matrices is minimal in the sense that there is a freedom to modify 
these $r$-matrices without changing the Poisson bracket \eqref{LLsp}. First of all, there is a trivial freedom of shifting  $r$ and $\mr$
as
\bea
\la{shift_sym}
r_{12}\to r_{12}+ f(\lambda/\mu)\mI\otimes \mI \, , ~~~\mr_{12}\to \mr_{12}+f(\lambda/\mu)\mI\otimes \mI\, ,
\eea
where $f$ is an arbitrary function  of the ratio of the spectral parameters. This redefinition affects neither the bracket (\ref{LL_trig}) nor
equations (\ref{shiftedCYB}), (\ref{shifted1}), (\ref{shifted2}). 

Second, one can redefine $\bar{r}$ and $r$ as
\bea
\la{non_triv_mod}
\begin{aligned}
&r(\lambda,\mu)\to r(\lambda,\mu)-s(\lambda)\otimes\mI+\mI\otimes s(\mu) \\
&\bar{r}(\lambda)\to \bar{r}(\lambda)-s(\lambda)\otimes \mI\, ,
\end{aligned}
\eea
where $s(\lambda)$ is an arbitrary matrix function of the spectral parameter. Owing to the structure of the 
bracket (\ref{sdL_t}) this redefinition of the $r$-matrices produces no effect on the latter, as $\mr$ remains unchanged, 
while the matrix $s$ decouples from the right-hand side of the $LL$ bracket (see \eqref{LLsp}).  For generic $s(\lambda)$,  redefinition (\ref{non_triv_mod}) affects\footnote{\la{footnote_sYB}An example of such a redefinition that does not 
affect the shifted Yang-Baxter equation corresponds to the choice $s(\lambda)=f(\lambda)\mI$, where $f$ is an arbitrary function of $\lambda$.}, however,  equations 
(\ref{shiftedCYB}), (\ref{shifted1}), (\ref{shifted2}). In particular, there exists a choice of $s(\lambda)$ which turns
the shifted Yang-Baxter equations for $\bar{r}$ and $r$ into the conventional ones, where the derivative terms on the right     
hand side of (\ref{shiftedCYB}), (\ref{shifted1}) and (\ref{shifted2}) are absent. One can take, for instance,
%\bea
%s(\lambda)=\frac{1}{\nn}\sum_{i\neq j}^{\nn}\Big(\frac{\gQ_i+\gQ_j}{2\gQ_{ij}}E_{ii}-\frac{\lambda\gQ_i-\gQ_j}{(\lambda-1)\gQ_{ij}}E_{ij}\Big)\, .
%\eea
\bea
s(\lambda)=\frac{1}{\nn}\sum_{i\neq j}^{\nn}\frac{\gQ_i}{\gQ_{ij}}(E_{ii}-E_{ij})+\frac{1}{\lambda-1}\frac{1}{N}\sum_{i\neq j}^{\nn}(E_{ii}-E_{ij})\, .
\eea
With the last choice the matrix $\bar{r}(\lambda)$ becomes 
\bea
\nonumber
\bar{r}(\lambda)
%&=&
%\sum_{i\neq j}\frac{\gQ_i}{\gQ_{ij}}(E_{ii}-E_{ij})\otimes \Big(E_{ii}-\frac{1}{\nn}\mI\Big)
%+\frac{1}{\lambda-1}\sum_{i\neq j}(E_{ii}-E_{ij})\otimes \Big(E_{jj}-\frac{1}{N}\mI\Big)\, \\
&=&\frac{1}{\lambda-1}\sum_{i\neq j}\frac{\lambda\gQ_i-\gQ_j}{\gQ_{ij}}(E_{ii}-E_{ij})\otimes \Big(E_{jj}-\frac{1}{\nn}\mI\Big)\, ,
\eea
while for $r(\lambda,\mu)$ one finds 
\bea
r_{12}(\lambda,\mu)=\frac{\lambda r^m_{12}+\mu r^m_{21}}{\lambda-\mu}+\frac{\rho_{12}}{\lambda-1} -\frac{\rho_{21}}{\mu-1} \, ,
\eea 
where 
$$
\rho_{12}=\sum_{i\neq j}(E_{ii}-E_{ij})\otimes \Big(E_{jj}-\frac{1}{\nn}\mI\Big)\, 
$$
and the modified $r$-matrix is 
\bea
\nonumber
r^m_{12}&=&
\sum_{i\neq j}^{\nn}\Big(\frac{\gQ_j}{\gQ_{ij}} E_{ii}-\frac{\gQ_i}{\gQ_{ij}} E_{ij}\Big)\otimes (E_{jj}-E_{ji})\\
\la{r_modified}
&-&\frac{1}{\nn}\sum_{i\neq j}\frac{\gQ_i}{\gQ_{ij}}(E_{ii}-E_{ij})\otimes \mI+\frac{1}{\nn}\sum_{i\neq j}\frac{\gQ_i}{\gQ_{ij}}\mI\otimes (E_{ii}-E_{ij})\, .
\eea
The modified $r$-matrix still solves the CYBE and obeys the same relation 
(\ref{prop_r}).
%$r_{12}^m+r_{21}^m=C_{12}-\mI\otimes \mI$.

There is no symmetry operating on $r$-matrices that would allow one to remove the scale-non-invariant terms from these matrices.    
Clearly, the $r$-matrices satisfying the shifted version of the Yang-Baxter equations have a simpler structure than their cousins 
subjected to the standard Yang-Baxter equations. This fact plays an important role when it comes to quantisation of the corresponding model and the 
associated algebraic structures. We also point out that the $r$-matrices we found here through considerations in appendix \ref{app:cauchy}
also follow from the elliptic  $r$-matrices of \cite{ACF} upon their hyperbolic degeneration, albeit modulo the shift symmetries (\ref{shift_sym}) 
and (\ref{non_triv_mod}). 

From (\ref{sdL_t}) one then finds 
\bea
\nonumber
\{{\rm Tr}_1L_1(\lambda),L_2(\mu)\}=[ {\rm Tr}_1L_1(\lambda)(r_{12}(\lambda,\mu)+\bar{r}_{21}(\mu)) , L_2(\mu)]\, ,
\eea
which, upon taking the limit $\lambda\to\infty$, yields  the Lax equation (\ref{evolLLpr}) with $M$ given by (\ref{M_k=1}).
The conserved quantities are, therefore,
$I_k(\lambda)={\rm Tr} L(\lambda)^k$, $k\in \mathbb{Z}$. The determinant 
$\det(L(\lambda)-\zeta\mI)$, 
which generates $I_k(\lambda)$ in the power series expansion over the parameter $\zeta$, defines 
the {\it classical spectral curve}
\bea
\det(L(\lambda)-\zeta\mI)=0\, , ~~~~~\zeta,\lambda\in {\mathbb C}\, .
\eea

\section{Quantum model}
\la{QMod}
\subsection{Quantum Heisenberg double}
At the classical level we obtained the  hyperbolic RS model by means of the Poisson reduction
of the Heisenberg double. It is therefore natural to start with the quantum analogue of the Heisenberg double. The Poisson algebra (\ref{AB}) can be straightforwardly quantised in the standard spirit
of deformation theory. We thus introduce an associative unital algebra generated by the entries of matrices $A,B$ modulo the relations \cite{SemenovTianShansky:1993ws}
\bea
\la{AB_quant}
\begin{aligned}
\RR_-^{-1}A_2 \RR_+A_1&=A_1 \RR_-^{-1}A_2 \RR_+ \, ,\\
\RR_-^{-1}B_2 \RR_+A_1&=A_1 \RR_-^{-1}B_2 \RR_- \, ,\\
\RR_+^{-1}A_2 \RR_+B_1&=B_1 \RR_-^{-1}A_2 \RR_+\, ,\\
\RR_-^{-1}B_2 \RR_+B_1&=B_1 \RR_-^{-1}B_2 \RR_+ \, ,
\end{aligned}
\eea
and they can be regarded as the quantisation of the Poisson relations (\ref{AB}). The quantum $\RR$-matrices here are defined as follows: first, we consider the following well-known solution of the quantum Yang-Baxter equation 
\bea
\la{Rstandard}
\RR=\sum_{i\neq j}^n E_{ii}\otimes E_{jj}+e^{\hbar/2}\sum_{i=1}^n E_{ii}\otimes E_{ii}
+ (e^{\hbar/2}-e^{-\hbar/2})\sum_{i>j}^n E_{ij}\otimes E_{ji}\, .
\eea
Using this $\RR$ one can construct two more solutions $\RR_{\pm}$ of the quantum Yang-Baxter equation, namely,  
\bea
\la{Rpm}
\RR_{+12}=\RR_{21}\, , ~~~~~\RR_{-12}=\RR_{12}^{-1}\, .
\eea
These solutions are, therefore, related as 
\bea
\la{relRplPmin}
\RR_{+21}\RR_{-12}=\mI\, ,
\eea
and they also satisfy
\bea
\la{PplminRmin}
\RR_+-\RR_-=(e^{\hbar/2}-e^{-\hbar/2})\,  C\, ,
\eea
where $C$ is the split Casimir. 
In the limit $\hbar\to 0$ the matrices $R_{\pm}$ expand as 
\bea
\RR_{\pm}=1+\hbar \rr_{\pm}+{\cal o}(\hbar)\, ,
\eea
where $\rr_{\pm}$ are the classical $\rr$-matrices (\ref{r-mat-gl}). Further, we point out that $\hat{\RR}_{\pm}=C \RR_{\pm}$ satisfy the {\it Hecke condition}
\bea
\la{Hecke_rel}
\widehat{\RR}_{\pm}^2 \mp (e^{\hbar/2}-e^{-\hbar/2})\hat{\RR}_{\pm}-\mI=\big(\hat{\RR}_{\pm} - e^{\pm\hbar/2}\mI\big)
\big(\hat{\RR}_{\pm} + e^{\mp\hbar/2}\mI\big)=0\, .
\eea

The first, or alternatively, the last line in (\ref{AB_quant})  is a set of defining relations for the corresponding subalgebra that describes a quantisation of the Semenov-Tian-Shansky bracket,
the latter has a set of Casimir functions generated by $C_k={\rm Tr}A^k$. In the quantum case an analogue ${\rm Tr}A^k$ can be defined by means of the quantum trace formula
$$
C_k={\rm Tr}_{\qq} A^k\,=\,{\rm Tr}(DA^k)\, , ~~~\qq \,=\, e^{-\hbar}\, , 
$$
where $D$ is a diagonal matrix $D={\rm diag}(\qq,\qq^2,\ldots , \qq^n)$. The elements $C_k$ are central in the subalgebra generated by $A$. Indeed, by successively using the permutation relations for $A$, one gets 
$$
A_2 \RR_+A_1^k \RR_+^{-1}=\RR_-A_1^k \RR_-^{-1}A_2 \, .
$$
We then multiply both sides of this relation by $D_1$ and take the trace in the first matrix space 
$$
A_2 {\rm Tr}_1\, (D_1 \RR_+A_1^k \RR_+^{-1})= {\rm Tr}_1\,(D_1 \RR_-A_1^k \RR_-^{-1})A_2\, .
$$
It remains to notice that $ {\rm Tr}_1\, (D_1 \RR_+A_1^k \RR_+^{-1})={\rm Tr}_1\,(D_1 \RR_-A_1^k \RR_-^{-1})={\rm Tr}_{\qq}A^k\cdot \mathbbm{1}$, so that 
\bea
A\, {\rm Tr}_{\qq}A^k= {\rm Tr}_{\qq}A^k\, A\, ,
\eea
{\it i.e.} ${\rm Tr}_{\qq}A^k$ is central in the subalgebra generated by $A$. Analogously, the $I_k={\rm Tr}_{\qq}B^k$ are central in the algebra generated by $B$ and, in particular, the $I_k$ form a commutative family. 

In principle, we can start with (\ref{AB_quant}) and try to develop a proper parametrisation of the $(A,B)$ generators suitable for reduction. 
It is an interesting path that should lead to understanding  how to implement the Dirac constraints at the quantum level. We will find, however, a short cut to the algebra of the quantum $L$-operator. 

\subsection{Quantum $R$-matrices and the $L$-operator}
An alternative route  to the quantum $R$-matrices and to the corresponding $L$-operator algebra is based on the observation that in the classical theory, the Poisson brackets between the entries
of the Lax matrix have the same structure (\ref{LL_trig}) for both rational and hyperbolic cases. As a consequence, the equations satisfied by the classical rational and hyperbolic $r$-matrices are also the same.  This should also be applied to the equations 
obeyed by the corresponding quantum $R$-matrices. We thus assume that the matrices $R$ and $\bar{R}$ for the hyperbolic RS model satisfy the system of equations 
\bea
\label{RRR}
R_{12}R_{13}R_{23}=R_{23}R_{13}R_{12} 
\eea
and 
\bea
\label{RRbRb}
R_{12}\bar{R}_{13}\bar{R}_{23}&=&
\bar{R}_{23}\bar{R}_{13}P_3R_{12}P_3^{-1},\\ 
\label{RbRb}
\bar{R}_{12}P_2\bar{R}_{13}P_2^{-1}&=&
\bar{R}_{13}P_3\bar{R}_{12}P_3^{-1}\, .
\eea   
and have the standard semi-classical limit where they match the classical $r$-matrices (\ref{r_trig}). Here and in the following \((Q_i,P_i)\) satisfy the quantum algebra
\begin{align}
Q_i Q_j\,=\,Q_j Q_i\qquad P_iP_j\,=\,P_jP_i \qquad [P_i ,Q_j]\,=\, (e^{\hbar}-1) Q_j P_j \delta_{ij}\, ,
\end{align}
being the standard quantisation of the Poisson algebra on the reduced phase space \eqref{QP},\eqref{PPred}.
In fact, it is not difficult to guess a proper solution for these $R$-matrices based on the analogy with the rational case. For $R$ we can take
\bea
\la{Rqh}
R=\exp{\hbar  r } \, ,
\eea
where $r$ is given on the first line of (\ref{r_trig}).
In the following we adopt  
the notation $R_+= R$.
Since the classical $r$-matrix satisfies the property $r^2=-r$, the exponential in  (\ref{Rqh}) can be easily evaluated and we find 
\bea
\la{Rqh_main}
R_+=\mI+(1-\qq)
\sum_{i\neq j}^{\nn}\Big(\frac{\gQ_j}{\gQ_{ij}} E_{ii}-\frac{\gQ_i}{\gQ_{ij}} E_{ij}\Big)\otimes (E_{jj}-E_{ji})\, .
\eea
A direct check shows that (\ref{Rqh_main}) is a solution of (\ref{RRR}).

In comparison to the rational model, a new feature is that 
there exists yet another solution $R_-$ of the Yang-Baxter equation, namely, 
\bea
\la{Rqh_main_min}
R_-=\mI-(1-\qq^{-1})
\sum_{i\neq j}^{\nn}(E_{ii}-E_{ij})\otimes \Big(\frac{\gQ_i}{\gQ_{ij}} E_{jj}-\frac{\gQ_j}{\gQ_{ij}} E_{ji}\Big)\, .
\eea
These solutions are related as 
\bea
\la{au_1} 
R_{+21}R_{-12}=\mI\, ,
\eea 
{\it i.e.} precisely in the same way as their non-dynamical counterparts, {\it cf.}
(\ref{relRplPmin}). Furthermore, the matrices  $R_{\pm}$ satisfy equation
\bea
R_+ - \qq R_{-}=(1-\qq)C\, .
\eea
They are also of Hecke type and the matrices $\hat{R}_{\pm}=CR_{\pm}$  have the following property
\bea
\big(\hat{R}_{\pm} - \mI\big)
\big(\hat{R}_{\pm} + \qq^{\pm 1}\mI\big)=0\, .
\eea

% Contrary to the rational case, this matrix does not satisfy the property (\ref{RR}),
% rather considering $R$ as a function of both $\hbar$ and $\gQ$, $R\equiv R(\hbar,\gQ)$, one has 
%\bea
%\nonumber
%R_{12}(\hbar,\gQ)R_{21}(-\hbar,1/\gQ)=\mI-(1-e^{-\hbar})\sum_{i,j=1}^{\nn} (E_{ii}\otimes E_{ji}-E_{ji}\otimes E_{ii})\, .
%\eea
%As a result, assuming for the corresponding  generators $T$ the algebraic
%relation $T_1T_2=T_2T_1R_{12}(\gQ)$ would  be incompatible. The same conclusion also follows from more complicated structure (\ref{TT_final}) of the Poisson bracket %between the components of $T$.  

Concerning the generalisation of equation (\ref{RRbRb}) to the hyperbolic case,  we can imagine two different versions - one involving $R_{+}$ and another $R_{-}$, that is,  
\bea
\la{RbarRpm}
%R_{\pm 12}(\gQ)\bar{R}_{13}(\gQ)\bar{R}_{23}(\gQ)=
%\bar{R}_{23}(\gQ)\bar{R}_{13}(\gQ)P_3R_{\pm 12}(\gQ)P_3^{-1},
R_{\pm 12}\bar{R}_{13}\bar{R}_{23}=
\bar{R}_{23}\bar{R}_{13}P_3R_{\pm 12}P_3^{-1},
\eea
It appears that there exists a unique matrix $\bar{R}$ which satisfies both these equations.  It is given by 
\bea
\la{bargh}
\bar{R}=\mI -\sum_{i\neq j}^{\nn}\frac{ \qq\gQ_i - \gQ_i}{\qq\gQ_i -  \gQ_j}(E_{ii} - E_{ij})\otimes  E_{jj}\, .
\eea
and its inverse is 
\bea
\la{bargh_inv}
\bar{R}^{-1}=\mI-(1-\qq)\sum_{i\neq j}^{\nn}\frac{ \gQ_i}{\gQ_{ij} }(E_{ii} - E_{ij})\otimes  E_{jj}\, .
\eea
The matrix (\ref{bargh})  also obeys (\ref{RbRb}), 
\bea
\la{barRbarR}
%\bar{R}_{12}(\gQ)P_2\bar{R}_{13}(\gQ)P_2^{-1}=
%\bar{R}_{13}(\gQ)P_3\bar{R}_{12}(\gQ)P_3^{-1}\, .
\bar{R}_{12}P_2\bar{R}_{13}P_2^{-1}=
\bar{R}_{13}P_3\bar{R}_{12}P_3^{-1}\, .
\eea
Introducing 
\bea
\la{defRGF}
\mR_{12}=\bar{R}_{12}^{-1} R_{12}\bar{R}_{21}\, ,
\eea
 we find
\bea
\la{hRNF}
\begin{aligned}
\mR_+&=\mI+(1-\qq)
\sum_{i\neq j}^{\nn}\frac{\gQ_i}{\gQ_{ij}}(E_{ij}\otimes E_{ji}-E_{ii}\otimes E_{jj})\, , \\
\mR_-&=\mI-(1-\qq^{-1})
\sum_{i\neq j}^{\nn}\frac{\gQ_j}{\gQ_{ij}}(E_{ij}\otimes E_{ji}-E_{ii}\otimes E_{jj})\, .
\end{aligned}
\eea
These matrices satisfy the Gervais-Neveu-Felder equation
\bea
\la{GNS_hyper}
%\mR_{\pm 12}(\gQ)P_2^{-1}\mR_{\pm 13}(\gQ)P_2\mR_{\pm 23}(\gQ)=P_1^{-1}\mR_{\pm 23}(\gQ)P_1\mR_{\pm 13}(\gQ)P_3^{-1}\mR_{\pm 12}(\gQ)P_3\, .
\mR_{\pm 12}P_2^{-1}\mR_{\pm 13}P_2\mR_{\pm 23}=P_1^{-1}\mR_{\pm 23}P_1\mR_{\pm 13}P_3^{-1}\mR_{\pm 12}P_3\, .
\eea
and are related to each other as
\bea
\la{au_2} 
 \mR_{+21}\mR_{-12}=\mI\, .
\eea
They also have another important property, usually referred to as the {\it zero weight condition} \cite{Felder:1994pb},
\bea
\la{zw}
[P_1P_2,\mR_{\pm}]=0\, .
\eea

Finally, the quantum $L$-operator is literally the same as its classical counterpart (\ref{Lax_red}), of course with the natural replacement of $p_i$ by the corresponding derivative
\bea
\la{Lax_red_hbar}
L=\sum_{i,j=1}^{\nn}\frac{\gQ_i -\omega \gQ_i}{\gQ_i - \omega \gQ_j }\, b_j \tt_j E_{ij}\, , ~~~~b_j\,=\, \prod\limits_{a\neq j}^{\nn}  \frac{ \omega \gQ_j -\gQ_a }{ \gQ_j- \gQ_a  }\, ,
\eea
where $\omega=e^{-\gamma}$ and $\tt_j$ is the operator $\tt_j=e^{-\hbar\frac{\pa}{\pa q_j}}$.\footnote{In fact, $\tt_j=P_j^{-1}$, we use $\tt_j$ to signify that we talk about a particular representation for $L$.}
On smooth functions $f(\gQ_1,\ldots, \gQ_{\nn})$ it acts as
$$
(\tt_j f)(\gQ_1,\ldots, \gQ_{\nn})=f(\gQ_1,\ldots, \qq \gQ_j\, , \ldots \gQ_{\nn})\, .
$$  
It is a straightforward exercise to check that this $L$-operator satisfies the algebraic relations 
\bea
\la{Rplmin}
\begin{aligned}
R_{+ 12}L_2\bar{R}^{-1}_{12}L_1&=L_1\bar{R}^{-1}_{21}L_2\mR_{+ 12}\, , \\
R_{- 12}L_2\bar{R}^{-1}_{12}L_1&=L_1\bar{R}^{-1}_{21}L_2\mR_{- 12}\, .
\end{aligned}
\eea
with the  $R$-matrices given by 
(\ref{Rqh_main}), (\ref{Rqh_main_min}), (\ref{bargh}) and (\ref{hRNF}). The consistency of these relations follow from (\ref{au_1}) and (\ref{au_2}). One can alternatively derive equations \eqref{Rplmin} by direct quantisation of \eqref{WW}-\eqref{WP}, where the classical matrix is chosen to be \(r_{12}\) or, equivalently, \(-r_{21}\)
\begin{align}
\label{WWq}
W_1 W_2 R_{\pm 12}\,&=\,R_{\pm 12} W_2 W_1\, , \\
W_1 \bar R_{12} P_2\,&=\,\bar R_{12} P_2 W_1\, ,
\end{align}
whose consistency follows from the same \(R\)-matrices relations. The algebraic relation \eqref{WWq} is also known as the \emph{quantum Frobenius group} condition \cite{Arutyunov:1996cmb}.
%It is convenient to introduce $R_{12}=R_{+21}$, so that $R_{-12}=R_{12}^{-1}$ and equations (\ref{Rplmin}) turn into
%\bea
%R_{12}=R_{+21}\, , ~~~~R_{-12}=R_{12}^{-1}\, .
%\eea
%Then equation above becomes 
%\bea
%R_{12}L_1\bar{R}_{21}^{-1}L_2\bar{R}_{12}^{-1}=L_2\bar{R}_{12}^{-1}L_1\bar{R}_{21}^{-1}R_{12}\, .
%\eea

Concerning commuting integrals, the Heisenberg double has a natural commutative family $I_k={\rm Tr}_{\qq}B^k$. It is not clear, however,
how these integrals can be expressed via $L$, because we are lacking an analogue of the quantum factorisation formula $B=TLT^{-1}$, where $T$ and $L$ 
would be subjected to well-defined algebraic relations. Instead, what we could do is to conjecture the same formula as was obtained for quantum integrals in the rational case \cite{Arutyunov:1996cmb}, where 
now the $R$-matrices are those of the hyperbolic model.   
Interestingly, the existence of two $R$-matrices, $R_{\pm}$, should give rise to two families of commuting integrals $I_k^{\pm}$. Borrowing 
the corresponding expression from the rational case  \cite{Arutyunov:1996cmb}, we conjecture the following {\it quantum trace formulae}
\bea
\la{qtr}
I_k^{\pm} = \text{Tr}_{12}\left(C_{12}^{\t_2} L_1 \bar{R}_{21}^{\t_2} R_{\pm12}^{\t_2} L_1 \dots L_1 \bar{R}_{21}^{\t_2} R_{\pm12}^{\t_2} L_1 \right)\, ,
\eea
as quantisation of the classical integrals (\ref{ctr}).
In (\ref{qtr}) the number $k$ on the right-hand side gives a number of $L_1$'s and $\t_2$ stands for the transposition in the second matrix space. 
In particular,
$$
C_{12}^{\t_2}=\sum_{i,j=1}^{N}E_{ij}\otimes E_{ij}\, 
$$
is a one-dimensional projector and from (\ref{Rqh_main}), (\ref{Rqh_main_min}) and (\ref{bargh}) we get 
\begin{align}
 \bar{R}_{21}^{\t_2} R_{+\,12}^{\t_2}=&\,\mI+ (1-\qq)\sum_{i,j} \frac{\gQ_i}{ \gQ_i- \qq \gQ_j}\,E_{ij}\otimes (E_{ij}-E_{jj}) \, ,
 \\
\bar{R}_{21}^{\t_2} R_{-\,12}^{\t_2}=&\, \mI + (1-\qq)\sum_{i,j} \left[\frac{\gQ_j}{\gQ_i-\qq \gQ_j}\,E_{ij}\otimes (E_{ij}- E_{jj}) +\frac{1}{\qq} (E_{ii}-E_{ij})\otimes E_{jj}\right]\, .
\end{align}
Commutativity of $I_k^{\pm}$ is then verified by tedious but direct computation which we do not reproduce here, rather our goal is to present a 
formula which relates $I_k^{\pm}$ with the commuting family given by Macdonald operators. 
  
We denote by $\{\mathcal{S}_k\}$ a commutative family of   {\it Macdonald operators}, where 
\bea
\la{Ruij_H_M}
\mathcal{S}_k&=&\omega^{\frac{1}{2}k(k-1)}\sum_{\substack{J\subset\{1,\ldots,n\} \\ |J|=k  }} \prod_{\substack{ i\in J \\ 
%a\in \{1,\ldots,n\}/J  
j\not\in J
} } \frac{\omega \gQ_i -\gQ_j }{ \gQ_i -\gQ_j } \prod_{i\in J} \tt_i \, .
\eea
The Macdonald operators have the following generating function
\bea
\la{gen_fun_Mac}
:\det (L-\zeta \mI):\, =\sum_{k=0}^{\nn} (-\zeta)^{\nn-k}\mathcal{S}_k\, , ~~~~\mathcal{S}_0=1\, ,
\eea
where  $\zeta$ is a formal parameter, $L$ is the Lax operator (\ref{Lax_red_hbar}). Under the sign $:\, :$ of normal ordering 
the operators $p_j$ and $q_j$ are considered as commuting and upon algebraic evaluation of the determinant all $\tt_j$ are 
brought to the right. In the classical theory the normal ordering is omitted and the corresponding generating function yields
classical integrals of motion that are nothing else but the spectral invariants of the Lax matrix.

%The meaning of the normal ordering here is the same as in (\ref{gen_fun_Mac_rat}).
We found an explicit formula that relates the families $\{I_k^{\pm}\}$ and $\mathcal{S}_k$. To present it, we need the notion of a $\qq$-number $[k]_{\qq}$
associated to an integer $k$
\bea
[k]_{\qq} = \sum_{n=0}^{k-1} \qq^n=\frac{~\, 1-\qq^k}{1-\qq}\, ,
\eea
so that $[k]_1=k$, which corresponds to the limit $\hbar\to 0$. Then ${\mathcal S}_k$ is expressed via $I_{m}^+$ or $I_{m}^-$ as 
\bea
\la{HI_Hyper}
{\mathcal S}_k=\frac{1}{[k!]_{\qq^{\pm 1}}}  \left| \begin{array}{ccccc}
I_1^{\pm} & ~~~[k-1]_{\qq^{\pm 1}}    & 0  & \cdots & 0 \\
I_2^{\pm} & I_1^{\pm} & ~~~[k-2]_{\qq^{\pm 1}}  & \cdots  & 0 \\
\vdots  & \vdots & \cdot & \cdots & \vdots \\
I_{k-1}^{\pm}  & I_{k-2}^{\pm} & \cdot & \cdots& ~~~[1]_{\qq^{\pm 1}} \\
I_k^{\pm} & I_{k-1}^{\pm} &  \cdot &  \cdots & I_1^{\pm} \end{array} \right| \, .
\eea
These formulae can be inverted  to express each integral \(I_k^{\pm}\) as the determinant of a \(k \times k \) matrix depending on \(\mathcal{S}_j\), namely,
\bea
I_{k}^{\pm}\,=\, 
\begin{vmatrix}
 \mathcal{S}_1 && 1 && 0 && \cdots && 0\\
 [2]_{\qq^{\pm 1}}\mathcal{S}_2 &&  \mathcal{S}_1 && 1 && 0 &&\cdots \\
 \vdots &&\vdots  && \cdots  && \cdots && 1\\
[k]_{\qq^{\pm 1}} \mathcal{S}_k &&  \mathcal{S}_{k-1}&&  \mathcal{S}_{k-2} && \cdots &&  \mathcal{S}_1\\
\end{vmatrix}
\, .
\eea

\subsection{Spectral parameter and quantum $L$-operator} 
The quantum $L$-operator depending on the spectral parameter is naturally introduced as a normal ordered version 
of its classical counterpart
\bea
\la{quantum_L_spect}
L(\lambda)=\frac{(1-\omega)}{\lambda}\sum_{i, j=1}^{\nn} \frac{\lambda \gQ_i-\omega e^{-\hbar/2}\gQ_j}{\gQ_i - \omega\gQ_j}b_j \tt_j E_{ij}=L-\frac{\omega \,e^{\hbar/2}}{\lambda}\gQ^{-1}L\gQ\, ,
\eea 
where $b_j$ are the same as in (\ref{Lax_red_hbar}).
This $L$-operator satisfies the following quadratic relation
\bea
\la{quadr_algebra_LL}
R_{12}(\lambda,\mu)L_2(\mu)\bar{R}_{12}^{-1}(\lambda)L_1(\lambda)=
L_1(\lambda)\bar{R}_{21}^{-1}(\mu)L_2(\mu)\mR_{12}(\lambda,\mu)\, ,
\eea
where 
\bea
\la{mR_spec}
\mR_{12}(\lambda,\mu)=\bar{R}_{12}^{-1}(\lambda)R_{12}(\lambda,\mu)\bar{R}_{21}(\mu)\, .
\eea
In (\ref{quadr_algebra_LL}) the quantum $R$-matrices are
\bea
\la{q_R_spec}
\begin{aligned}
R(\lambda,\mu)&=\frac{\lambda e^{\hbar/2}R_+-\mu e^{-\hbar/2}R_-}{\lambda-\mu}-\frac{e^{\hbar/2}-e^{-\hbar/2}}{e^{\hbar/2}\lambda-1}X_{12}
+\frac{e^{\hbar/2}-e^{-\hbar/2}}{e^{-\hbar/2}\mu -1}X_{21} \, . \\
\bar{R}(\lambda)&=\bar{R}-\frac{e^{\hbar}-1}{e^{\hbar/2}\lambda-1} X_{12}\, .
\end{aligned}
\eea
Here $R_+$ and $R_-$ are the solutions (\ref{Rqh_main}) and (\ref{Rqh_main_min}) of the quantum Yang-Baxter equation, $\bar{R}$ is (\ref{bargh})
and we have introduced the matrix $X\equiv X_{12}$,
\bea
\la{Xmat}
X=\sum_{i,j=1}^N E_{ij}\otimes E_{jj}\, .
%=\mI\otimes \mI-\sigma_{12}\, .
\eea
This matrix satisfies a number of simple relations with $\bar{R}$ and $R_{\pm}$, which are
\bea
\bar{R} X=X\bar{R}
\eea
and 
\bea
\la{RpmX}
\begin{aligned}
R_-X_{12}&= X_{12} R_-\, , ~~~~~~  R_-X_{21}-X_{21}R_-=(1-\qq^{-1})(X_{12}-X_{21})\, , \\
R_+X_{21}&=X_{21}R_+\, , ~~~~~~  R_+X_{12}-X_{12}R_+ = -(1-\qq)(X_{12}-X_{21})\, .
\end{aligned}
\eea
We also present the formula for the inverse of $\bar{R}(\lambda)$
\bea
\bar{R}(\lambda)^{-1}=\bar{R}^{-1}+\frac{e^{\hbar}-1}{e^{\hbar/2}\lambda-e^{\hbar}}X_{12}\, .
\eea
With the help of this formula and (\ref{q_R_spec}) one can show that (\ref{mR_spec}) boils down to 
\bea
\la{unitarity_mR}
\mR_{12}(\lambda,\mu)=\frac{\lambda e^{\hbar/2}\mR_+-\mu e^{-\hbar/2}\mR_-}{\lambda-\mu}\, ,
\eea
where $\mR_{\pm}$ are the same as given by  (\ref{hRNF}). We note also the relation 
\bea
\hspace{-0.5cm}
R_{12}(\lambda,\mu)R_{21}(\mu,\lambda)=\mR_{12}(\lambda,\mu)\mR_{21}(\mu,\lambda)=\frac{(e^{\hbar/2}\lambda-e^{-\hbar/2}\mu)(e^{-\hbar/2}\lambda-e^{\hbar/2}\mu)  }{(\lambda-\mu)^2}\mI\, .
\eea
Finally, in addition to (\ref{mR_spec}) there is one more relation between $R(\lambda,\mu)$ and $\mR(\lambda,\mu)$, namely,
\bea
\la{an_id}
\mR_{12}(\lambda,\mu)=P_1^{-1}\bar{R}_{21}(\mu)P_1 R_{12}(\lambda,\mu) P_2^{-1}\bar{R}_{12}^{-1}(\lambda)P_2\, .
\eea

An interesting observation is that the combination 
\bea
\nonumber
R^{\rm YB}(\lambda,\mu)=\frac{\lambda e^{\hbar/2}R_+-\mu e^{-\hbar/2}R_-}{\lambda-\mu}
\eea
solves the usual quantum Yang-Baxter equation with the spectral parameter. However, the full $R$-matrix
in (\ref{q_R_spec}) differs from  $R^{\rm YB}$ by the terms that violate scale invariance. As a result, this matrix obeys the 
shifted version of the quantum Yang-Baxter equation, namely, 
\bea
\la{qs1}
R_{12}(\lambda,\mu)R_{13}(\qq \lambda,\qq \tau)R_{23}(\mu,\tau)=R_{23}(\qq \mu,\qq \tau)R_{13}(\lambda,\tau)R_{12}(\qq\lambda,\qq\mu)\, .
\eea
In addition, there are two more equations -- the one involving both $R$ and $\bar{R}$, and the other involving $\bar{R}$ only,
\bea
\la{qs2}
R_{12}(\lambda,\mu)\bar{R}_{13}(\qq \lambda)\bar{R}_{23}(\mu)&=&
\bar{R}_{23}(\qq \mu)\bar{R}_{13}(\lambda)P_3 R_{12}(\qq\lambda,\qq\mu)P_3^{-1}\, , 
\eea
\bea
\la{qs3}
\bar{R}_{12}(\lambda)P_2\bar{R}_{13}(\qq\lambda)P_2^{-1}&=&\bar{R}_{13}(\lambda)P_3\bar{R}_{12}(\qq\lambda)P_3^{-1}\, .
\eea
It is immediately recognisable that equations (\ref{qs1}), (\ref{qs2}) and (\ref{qs3}) are a quantum analogue (quantisation) of 
the classical equations (\ref{shiftedCYB}), (\ref{shifted1})  and (\ref{shifted2}), respectively. 
In the semi-classical expansion  
\bea
R(\lambda,\mu)=\mI+\hbar r(\lambda,\mu)+o(\hbar)\, , ~~~
\bar{R}(\lambda)=\mI+\hbar r(\lambda)+o(\hbar)\, 
\eea 
the matrices (\ref{q_R_spec}) yield 
\bea
%\la{r_spec_semi}
\nonumber
\begin{aligned}
r_{12}(\lambda,\mu)&=\frac{\lambda r_{12}+\mu r_{21}}{\lambda-\mu}+\frac{\sigma_{12}}{\lambda-1} -\frac{\sigma_{21}}{\mu-1}\\
&~~~~~~~~~+\Big(\frac{1}{2}\frac{\lambda+\mu}{\lambda-\mu}-\frac{1}{\lambda-1}+\frac{1}{\mu-1}\Big)\mI\otimes \mI \, ,\\
\bar{r}_{12}(\lambda)&=\bar{r}_{12}+\frac{\sigma_{12}}{\lambda-1}-\frac{\mI\otimes \mI}{\lambda-1} \, , 
\end{aligned}
\eea
which is different from the canonical classical $r$-matrices (\ref{spec_dep_r_final}) by allowed symmetry shifts.
Thus, (\ref{q_R_spec}) should be regarded  as a quantisation of the classical $r$-matrices satisfying the shifted Yang-Baxter 
equation. In this respect it is interesting to point out that the corresponding quantisation of the $r$-matrices 
solving the usual CYBE remains unknown. 

Finally, the algebra  (\ref{quadr_algebra_LL}) should be completed by the following additional relations encoding the commutation properties 
of $L$ with $\gQ$  
\bea
%P_1^{-1}\gQ_2&=&\gQ_2P_1^{-1}\Omega_{12} \, ~~~\gQ_2^{-1}P_1=P_1\gQ_2^{-1}\Omega_{12}^{-1}\, ,\\
L_1\gQ_2&=&\gQ_2L_1\Omega_{12}\, , ~~~\gQ_1^{-1}L_2=L_2\gQ_1^{-1}\Omega_{12}\, , 
\eea
where $\Omega_{12}=\mI-(1-\qq)\bar{C}_{12}$.

Now we derive a couple of important consequences of the algebraic relation (\ref{quadr_algebra_LL}). Namely, we establish the quantum Lax representation, similar to the rational case, and also prove the commutativity of the operators ${\rm Tr}L(\lambda)$ for different values of the spectral parameter. 

Following considerations
of the dynamics in the classical  theory, we take $H=\lim\limits_{\lambda\to\infty} {\rm Tr} L(\lambda)$ as the hamiltonian.
From (\ref{quadr_algebra_LL}) we get 
\bea
\la{quadr_algebra_LL_Tr}
{\rm Tr}_1\Big[R_{21}(\mu,\lambda)L_1(\lambda)\bar{R}_{21}^{-1}(\mu)\Big]\, L_2(\mu)=
L_2(\mu)\, {\rm Tr}_1\Big[\bar{R}_{12}^{-1}(\lambda)L_1(\lambda)\mR_{21}(\mu,\lambda)\Big] \, ,
\eea
where (\ref{unitarity_mR}) was used. A straightforward computation reveals that the traces on the left and the right-hand side of the last expression 
are equal and that, for instance, 
\bea
e^{\hbar/2}{\rm Tr}_1\Big[\bar{R}_{12}^{-1}(\lambda)L_1(\lambda)\mR_{21}(\mu,\lambda)\Big]={\rm Tr}L(\lambda) \, \mI - 
M(\lambda,\mu)\, ,
\eea
where 
\bea
\la{quantum_M}
\begin{aligned}
M(\lambda,\mu)&= (e^{\hbar}-1)\frac{\lambda}{\lambda-\mu}\frac{\mu-e^{-\hbar/2}}{\lambda-e^{\hbar/2}}\, L(\lambda)
 \\
&+ \frac{e^{\hbar}-1}{\lambda-e^{\hbar/2}}
\sum_{i\neq j}^{\nn}\frac{\lambda e^{-\hbar}\gQ_j-e^{-\hbar/2}\gQ_i}{\gQ_i-e^{-\hbar}\gQ_j}
L_{ij}(\lambda)(E_{ii}-E_{ij})\, .
\end{aligned}
\eea
Thus, equation (\ref{quadr_algebra_LL_Tr}) turns into 
\bea
\la{q_LE}
{\rm Tr}L(\lambda)L(\mu)-L(\mu){\rm Tr}L(\lambda)=[M(\lambda,\mu),L(\mu)]\, .
\eea
From (\ref{quantum_M}) we, therefore, derive the quantum-mechanical operator $M$
\bea
\nonumber
M=\lim_{\lambda\to\infty} M(\lambda,\mu)&=&(e^{\hbar}-1)\sum_{i\neq j}^{\nn}\frac{e^{-\hbar}\gQ_j}{\gQ_i-e^{-\hbar}\gQ_j}
L_{ij}(E_{ii}-E_{ij})\\
&&\hspace{1cm} =(e^{\hbar}-1)\sum_{i\neq j}^{\nn}L_{ij}\frac{\gQ_j}{\gQ_{ij}}
(E_{ii}-E_{ij})\, ,
\eea
where in the last expression we commuted the entries of $L_{ij}$ to the left so that it formally coincides with its classical counterpart 
(\ref{M_k=1}). In the limit $\lambda\to \infty$, (\ref{q_LE}) becomes the quantum Lax equation. Note that in the derivation of this equation we did not 
use any concrete form of $L$; we only use that it factorises as $L=WP^{-1}$, where $W$ is a function of coordinates only. 

Taking the trace of (\ref{q_LE}), one gets 
\bea
\la{q_LE_1}
{\rm Tr}L(\lambda){\rm Tr}L(\mu)-{\rm Tr}L(\mu){\rm Tr}L(\lambda)={\rm Tr} [M(\lambda,\mu),L(\mu)]\, .
\eea
A priori the trace of the commutator on the right-hand side might not be equal to zero, because it involves matrices with operator-valued entries. 
An involved calculation that uses representation (\ref{quantum_L_spect}) shows that it nevertheless vanishes\footnote{For this result to hold, the presence in (\ref{quantum_M}) of the first term proportional to $L(\lambda)$ is of crucial importance.}, identically for 
$\lambda$ and $\mu$. Fortunately, there is a simple and transparent way to show the commutativity of traces of the Lax operator, which directly relies 
on the algebraic relations (\ref{quadr_algebra_LL_Tr}), thus bypassing the construction of the quantum Lax pair. 
Indeed, let us multiply both sides of (\ref{quadr_algebra_LL}) with $P_2^{-1}\bar{R}_{12}(\lambda)P_2 R_{12}^{-1}(\lambda,\mu)$ and take the trace 
with respect to both spaces. We get 
\bea
\nonumber
&&{\rm Tr}_{12}\Big[P_2^{-1}\bar{R}_{12}(\lambda)P_2L_2(\mu)\bar{R}_{12}^{-1}(\lambda)L_1(\lambda)\Big]=\\
\nonumber
&&\hspace{2cm}
{\rm Tr}_{12}\Big[ P_2^{-1}\bar{R}_{12}(\lambda)P_2 R_{12}^{-1}(\lambda,\mu)L_1(\lambda)\bar{R}_{21}^{-1}(\mu)L_2(\mu)\mR_{12}(\lambda,\mu)\Big]\, .
\eea 
From (\ref{an_id}) we have 
\bea
\nonumber
P_2^{-1}\bar{R}_{12}(\lambda)P_2 R_{12}^{-1}(\lambda,\mu)=\mR_{12}^{-1}(\lambda,\mu)P_1^{-1}\bar{R}_{21}(\mu) P_1\, ,
\eea
so that the right-hand side of the above equation can be transformed as 
\bea
\la{trace_com}
&&{\rm Tr}_{12}\Big[P_2^{-1}\bar{R}_{12}(\lambda)P_2L_2(\mu)\bar{R}_{12}^{-1}(\lambda)L_1(\lambda)\Big]=\\
\nonumber
&&\hspace{2cm}
{\rm Tr}_{12}\Big[\mR_{12}^{-1}(\lambda,\mu)P_1^{-1}\bar{R}_{21}(\mu) P_1L_1(\lambda)\bar{R}_{21}^{-1}(\mu)L_2(\mu)\mR_{12}(\lambda,\mu)\Big]\, .
\eea 
Further progress is based on the fact that the matrices $\bar{R}_{12}(\lambda)$ and $\bar{R}_{12}^{-1}(\lambda)$  are diagonal in the second space. 
We represent it in factorised form 
\bea
\bar{R}_{12}(\lambda)=\sum_{j=1}^{\nn} G_j(\lambda)\otimes E_{jj}\, , 
\eea
see (\ref{q_R_spec}), (\ref{bargh}) and (\ref{Xmat}). 
Therefore, 
\bea
\la{shift_bar}
P_2^{-1}\bar{R}_{12}(\lambda)P_2=\sum_{j=1}^{\nn} P_j^{-1}G_j(\lambda)P_j\otimes E_{jj}\, .
\eea
Although this expression involves the shift operator, it commutes with any function of coordinates $q_j$, because when pushed through (\ref{shift_bar}),
this function will undergo the shifts of $q_j$ in opposite directions  which compensate each other.
Similarly, 
\bea
\nonumber
\bar{R}_{12}^{-1}(\lambda)=\sum_{j=1}^{\nn} G_j(\lambda)^{-1}\otimes E_{jj}=\sum_{j=1}^{\nn}(\mI\otimes E_{jj})(G_j(\lambda)^{-1} \otimes\mI)\, .
\eea
 Consider first the left-hand side of (\ref{trace_com})
\bea
\nonumber
{\rm Tr}_{12}\Big[\sum_{j=1}^{\nn} \sum_{k=1}^{\nn} (P_j^{-1}G_j(\lambda)P_j\otimes E_{jj}L(\mu)E_{kk} )(G_k(\lambda)^{-1}\otimes\mI)
L_1(\lambda)\Big]\, .
\eea
Using the cyclic property of the trace in the second space, this expression is equivalent to 
\bea
\nonumber
{\rm Tr}_{12}\Big[\sum_{j=1}^{\nn} \sum_{k=1}^{\nn} (P_j^{-1}G_j(\lambda)P_j\otimes L(\mu)E_{jj}E_{kk} )(G_k(\lambda)^{-1}\otimes\mI)
L_1(\lambda)\Big]\, .
\eea
Taking into account that $L=WP^{-1}$ and the commutativity of $P_j^{-1}G_j(\lambda)P_j$ with any function of coordinates, we arrive at 
\bea
\nonumber
{\rm Tr}_{12}\Big[\sum_{j=1}^{\nn} (\mI\otimes W(\mu))(P_j^{-1}G_j(\lambda)P_j\otimes P^{-1}_jE_{jj})\bar{R}_{12}^{-1}(\lambda)
L_1(\lambda)\Big]=
%{\rm Tr}_{12}\Big[\sum_{j=1}^{\nn} (\mI\otimes L(\mu))(G_j(\lambda)\otimes E_{jj})\bar{R}_{12}^{-1}(\lambda)
%L_1(\lambda)\Big]
{\rm Tr}L(\mu) {\rm Tr}L(\lambda)
\, .
\eea
Now we look at the right-hand side of (\ref{trace_com}): using the cyclic property of the trace, the matrix $\mR_{12}(\lambda,\mu)$
can be moved to the left where it cancels with its inverse. This manipulation is allowed because $L_1(\lambda)$ and $L_2(\mu)$
produce together a factor $P_1^{-1}P_2^{-1}$ with which $\mR_{12}(\lambda,\mu)$ commutes due to the zero weight condition (\ref{zw}).
Also, the individual entries of $\mR_{12}(\lambda,\mu)$ are freely moved through $P_1^{-1}\bar{R}_{21}(\mu) P_1$, because of the diagonal structure of the latter matrix  in the first matrix space, analogous 
to the similar property of (\ref{shift_bar}). Then, to eliminate $\bar{R}_{21}(\mu)$, one employs  the same procedure as was used  
for the left-hand side of (\ref{trace_com}) and the final result is ${\rm Tr}L(\lambda) {\rm Tr}L(\mu)$.
This proves the commutativity of traces of the Lax matrix for different values of the spectral parameter. 

We finally remark that writing the analogue of \eqref{gen_fun_Mac} with spectral parameter dependent Lax operator \cite{Has,Antonov:1997zc}
\bea
:\det (L(\lambda
)-\zeta \mI):\, =\sum_{k=0}^{\nn} (-\zeta)^{\nn-k}\mathcal{S}_k(\lambda
)\, ,
\eea
the quantities \(\mathcal{S}_k(\lambda
)\) are commuting integrals and they are related to Macdonald operators \eqref{Ruij_H_M} by a simple coupling- and spectral parameter-dependent rescaling
\begin{align}
\mathcal{S}_k(\lambda) \,=\, \lambda^{-k}(\lambda -\omega^k\, e^{-\hbar/2})(\lambda-e^{{-\hbar}/{2}})^{k-1}\,\mathcal{S}_k\, .
\end{align}

\section{Conclusions}

We have discussed the hyperbolic RS model in the context of Poisson reduction of the Heisenberg double \cite{Arutyunov:1996uw}: we derive its Poisson structure and show that only on the reduced phase space does the Poisson algebra of the Lax matrix close and take a form very similar to the Lax matrix of the rational RS model \cite{Arutyunov:1996cmb}. We find a quantisation of the $L$-operator algebra governed by new $R$ matrices $R_{\pm}$, along with a quantisation of the classical integrals in the form of quantum trace formulae $I_k$ (see (\ref{qtr})). We show how these quantum integrals are related to the well-known Macdonald operators through determinant formulae. Along the way we present a second Lax matrix that we can use to introduce a spectral parameter in the model. At the classical level this yields $r$-matrices that satisfy the shifted Yang-Baxter equation due to scale-violating terms. We show that this $L$-operator algebra admits a quantisation as well, with new $R$ matrices satisfying the shifted quantum Yang-Baxter equation. 

A particularly interesting observation is that one cannot obtain the quantum $L$-operator algebra from the quantum Heisenberg double in the same way as was done for the quantum cotangent bundle. It would be interesting to pursue the question whether and how one can impose the Dirac constraints after quantisation in order to reconstruct the quantum $L$-operator algebra. 
A first step in that direction could be finding an analytic proof that the Dirac bracket for $L$ on the reduced phase space is closed for general $N$. Another interesting question is to find the relation between our quantum trace formulae and the commuting traces obtained by the fusion procedure \cite{Avan:2003ke,Nagy:2004jv} for the equations \eqref{Rplmin}. 
In addition, it would be interesting to extend our results to the RS models with spin, in particular, to those discussed in \cite{Chalykh:2018wce}, as well as to find an analogue of the formulae \eqref{qtr} for the model with elliptic potential or for other series of Lie algebras. Constructing the quantum spin versions of these models could further aid the understanding of the RS type models that appear in the study of conformal blocks as in \cite{Isachenkov:2017qgn}.

\section*{Acknowledgements} 
We would like to thank Sylvain Lacroix for interesting discussions. 
The work of G. A. and E. O. is funded by the Deutsche Forschungsgemeinschaft (DFG, German Research 
Foundation) under Germany's Excellence Strategy -- EXC 2121 ``Quantum Universe" -- 390833306.
The work of E.O. is also supported by the DFG under the Research Training Group 1670.

%%%%%%%%%%%%%%%%%%%%%
\appendix
\section{Derivation of the Poisson structure}
\subsection{Lax matrix and its Poisson structure}
Consider the following matrix function on the Heisenberg double
\bea
\la{L_on_HD}
L=T^{-1}BT\, ,
\eea
where $T$ is the Frobenius solution of the factorisation problem (\ref{Trig_fact}). On the reduced space $L$ turns into the Lax matrix of the hyperbolic RS model. 
For this reason we continue to call (\ref{L_on_HD}) the Lax matrix and below we compute the Poisson brackets between the entries of $L$ considered as functions on the Heisenberg double. This will constitute the first step
towards evaluation of the corresponding Dirac bracket.  

The standard manipulations give 
\bea
\la{LL_ap}
\{L_1,L_2\}&=&\mathbb{T}_{12}L_1L_2-L_1 \mathbb{T}_{12}L_2-L_2\mathbb{T}_{12}L_1+L_1L_2\mathbb{T}_{12}\\
&+&T_1^{-1}T_2^{-1}\{B_1B_2\}T_1T_2\nonumber
+\mathbb{B}_{21} L_2
-L_2 \mathbb{B}_{21} -\mathbb{B}_{12} L_1
+L_1 \mathbb{B}_{12}\, , 
\eea
where we defined the following quantities 
\bea
\nonumber
\mathbb{T}_{12}&=&T_1^{-1}T_2^{-1}\{T_1,T_2\} \, ,\\
\nonumber
\mathbb{B}_{12}&=&T_1^{-1}T_2^{-1}\{T_1,B_2\}T_2\, .
%\mathbb{W}_{21}&=&T_1^{-1}T_2^{-1}\{T_2,B_1\}T_1
\eea
By using (\ref{AB}), we get 
\bea
 \nonumber
T_1^{-1}T_2^{-1} \{B_1,B_2\}T_1T_2&=& -\rdr_-\, L_1L_2 - L_1L_2 \rdr_+ + L_1\rdr_-L_2 +L_2\rdr_+ L_1\, .
 \eea
 Here we introduced the {\it dressed} $\rr$-matrices
 \bea
 \rdr_{\pm}=T_1^{-1}T^{-1}_2\rr_{\pm}T_1T_2\, , 
 \eea
 which have proved themselves to be a useful tool for the present calculation. 
 The dressed $\rr$-matrices have essentially the same properties as their undressed counterparts, most importantly, 
 \bea
 \la{rdrid}
  \rdr_{+}- \rdr_{-}=C_{12}\, ,
 \eea
 because $C_{12}$ is an invariant element. 
 Thus, for (\ref{LL_ap}) we get
 \bea
 \nonumber
\{L_1,L_2\}&=&(\mathbb{T}_{12} -\rdr_-)L_1L_2+L_1L_2(\mathbb{T}_{12}-\rdr_+) +L_1 (\rdr_--\mathbb{T}_{12})L_2+L_2(\rdr_+-\mathbb{T}_{12})L_1\\
\la{LL_ap1}
&+&\mathbb{B}_{21} L_2 -L_2 \mathbb{B}_{21} -\mathbb{B}_{12} L_1+L_1 \mathbb{B}_{12}\, .
\eea

Now we proceed with evaluation of $\mathbb{T}$. Taking onto account that $T$ satisfies (\ref{Frob}), in components we have 
\bea
\la{TTap}
\mathbb{T}_{ij,kl}&=&T^{-1}_{ip}T^{-1}_{kq}\frac{\delta T_{pj}}{\delta A_{mn}}\frac{\delta T_{ql}}{\delta A_{rs}}\{A_{mn},A_{rs}\}\\
\nonumber
%\hspace{-0.5cm}
&=&
\sum_{a\neq j}\sum_{b\neq l}\frac{1}{\gQ_{ja}\gQ_{lb}}(\delta_{ia}T_{nj}T_{am}^{-1}+\delta_{ij}T_{na}T_{jm}^{-1})(\delta_{kb}T_{sl}T_{br}^{-1}+\delta_{kl}T_{sb}T_{lr}^{-1})\{A_{mn},A_{rs}\}\, \\
\nonumber
&=&\sum_{a\neq j}\sum_{b\neq l}\frac{1}{\gQ_{ja}\gQ_{lb}}(\delta_{ia}\delta_{kb} \,  \zeta_{aj,bl} +\delta_{ij}\delta_{kb}\, \zeta_{ja,bl} 
+\delta_{ia}\delta_{kl}\, \zeta_{aj,lb} +\delta_{ij}\delta_{kl}\, \zeta_{ja,lb})\, .
\eea
Here $\gQ_{ij}=\gQ_i-\gQ_j$ and we introduced the concise notation
\bea
\nonumber
\zeta_{12}=T_1^{-1}T_2^{-1}\{A_1,A_2\}T_1T_2\, .
\eea
Using (\ref{AB}) and the fact that $A=T\gQ T^{-1}$, we find that  
\bea
\nonumber
\zeta_{12}=-\rdr_-\, \gQ_1\gQ_2-\gQ_1\gQ_2\, \rdr_+    +   \gQ_1\, \rdr_-\gQ_2 + \gQ_2\, \rdr_+\gQ_1\, .
\eea
With the help of (\ref{rdrid}) we find in components 
\bea
\nonumber 
\zeta_{ij, kl}&=&-\gQ_{ij}(\rdr_{- ij, kl} \gQ_{kl} +C_{ij,kl}\gQ_k)\, ,
\eea
where $C_{ij,kl}=\delta_{jk}\delta_{il}$ are the entries of $C_{12}$. Substitution of this tensor into (\ref{TTap}) yields the following expression 
\bea
\nonumber
\mathbb{T}_{ij,kl}&=&  \sum_{a\neq j}\sum_{b\neq l}\Big(                                                                                                                                                                          
-\delta_{ia}\delta_{kb} \rdr_{- aj, bl}   +\delta_{ij}\delta_{kb} \rdr_{- ja, bl} 
+\delta_{ia}\delta_{kl} \rdr_{- aj, lb}   -\delta_{ij}\delta_{kl} \rdr_{- ja, lb}  \Big) \\
\nonumber
&+&\sum_{a\neq j}\sum_{b\neq l}\frac{1}{\gQ_{lb}}\Big(                                                                                                                
\delta_{ia}\delta_{kb} C_{aj,bl}\gQ_b   -\delta_{ij}\delta_{kb} C_{ja,bl}\gQ_b + \delta_{ia}\delta_{kl} C_{aj,lb}\gQ_l - \delta_{ij}\delta_{kl} C_{ja,lb}\gQ_l \Big)\, .
\eea
In the first line the summation can be extended to all values of $a$ and $b$, because the expression which is summed vanishes for $a=j$ and independently for $b=l$.
For the same reason, we have extended the summation over $a$ in the second line, where we also substitute the explicit value for $C_{ij,kl}=\delta_{jk}\delta_{il}$.
In this way we find 
\bea
\nonumber
\mathbb{T}_{ij,kl}&=&  \sum_{ab}\Big(                                                                                                                                                                          
-\delta_{ia}\delta_{kb} \rdr_{- aj, bl}   +\delta_{ij}\delta_{kb} \rdr_{- ja, bl} 
+\delta_{ia}\delta_{kl} \rdr_{- aj, lb}   -\delta_{ij}\delta_{kl} \rdr_{- ja, lb}  \Big) \\
\nonumber
&+&\sum_{a }\sum_{b\neq l}\frac{1}{\gQ_{lb}}\Big(                                                                                                                
\delta_{ia}\delta_{kb}\delta_{al}\delta_{jb} \gQ_b   -\delta_{ij}\delta_{jl}\delta_{kb}\delta_{ab}\gQ_b + \delta_{ia}\delta_{kl}\delta_{ab}\delta_{jl}\gQ_l - \delta_{ij}\delta_{kl}\delta_{al}\delta_{jb} \gQ_l \Big)\, .
\eea
This further yields the following expression 
\bea
\nonumber
\mathbb{T}_{ij,kl}&=& - \rdr_{- ij, kl}   +\delta_{ij}\sum_a  \rdr_{- ia, kl} 
+\delta_{kl}\sum_a \rdr_{- ij, k a}   -\delta_{ij}\delta_{kl} \sum_{ab}\rdr_{- ia, kb} \\
\nonumber
&+& \sum_{b\neq l}\frac{1}{\gQ_{lb}}\Big(                                                                                                                
\delta_{kb}\delta_{il}\delta_{jb} \gQ_b   -\delta_{ij}\delta_{jl}\delta_{kb}\gQ_b +\delta_{kl}\delta_{ib}\delta_{jl}\gQ_l - \delta_{ij}\delta_{kl}\delta_{jb} \gQ_l \Big)\, .
\eea
Here the second line can be written in the concise form as the matrix element $r_{\gQ ij,kl}$ of the following matrix 
\bea
\la{rgQ}
r_{\gQ}=\sum_{a\neq b}\frac{\gQ_b}{\gQ_{ab}} (E_{aa}-E_{ab})\otimes (E_{bb}-E_{ba})\, 
\eea
Therefore,
\bea
\nonumber
\mathbb{T}_{ij,kl}&=&r_{\gQ  ij,kl} - \rdr_{- ij, kl}   +\delta_{ij}\sum_a  \rdr_{- ia, kl} 
+\delta_{kl}\sum_a \rdr_{- ij, k a}   -\delta_{ij}\delta_{kl} \sum_{ab}\rdr_{- ia, kb} \, .
\eea
Hence, 
\bea
\la{TT_final}
\mathbb{T}_{12}=r_{\gQ  12}-\rdr_{-12}+a_{12}+b_{12}-c_{12}\, .
\eea
where we introduced three $r$-matrices, $a$, $b$ and $c$ with entries 
\bea
a_{ij,kl}=\delta_{ij}\sum_a  \rdr_{- ia, kl} \, , ~~~b_{ij,kl}=\delta_{kl}\sum_a \rdr_{- ij, k a}\, , ~~~c_{ij,kl}=\delta_{ij}\delta_{kl} \sum_{ab}\rdr_{- ia, kb}\, .
\eea
Needless to say, the bracket thus obtained is compatible with the Frobenius condition (\ref{Frob}), which means that 
$$
\sum_a\mathbb{T}_{ia,kl}=0 \, , ~~~\sum_a\mathbb{T}_{ij,ka}=0\, ,
$$ 
for any values of the free indices. 

Now we turn our attention to $\mathbb{B}_{12}$, which in components reads as 
\bea
\nonumber
\mathbb{B}_{ij,kl}&=&\sum_{a\neq j}\frac{1}{\gQ_{ja}}(\delta_{ia}\eta_{aj,kl}+\delta_{ij}\eta_{ja,kl})\, ,
\eea 
where we introduced the notation 
\bea
\nonumber
\eta_{12}=T^{-1}_1T^{-1}_2\{A_1,B_2\}T_1T_2\, .
\eea
With the help of (\ref{AB}) we get
 \bea
 \nonumber
 \eta_{12}=-\rdr_-\, \gQ_1L_2-\gQ_1L_2\, \rdr_-   +  \gQ_1\, \rdr_-L_2+L_2\, \rdr_+\gQ_1\, , 
 \eea
 and by using (\ref{rdrid}) obtain for components the following expression 
 \bea
 \nonumber
\eta_{aj,kl}&=&\gQ_{ja}(L_{ks}\rdr_{-aj,sl}-\rdr_{-aj,ks}L_{sl})+L_{ks}C_{aj,sl}\gQ_j\, .
 \eea
With this expression at hand, we get 
\bea
\nonumber
\mathbb{B}_{ij,kl}&=&\sum_{a\neq j}\Big(\delta_{ia}(L_{ks}\rdr_{-aj,sl}-\rdr_{-aj,ks}L_{sl}) -\delta_{ij} (L_{ks}\rdr_{-ja,sl}-\rdr_{-ja,ks}L_{sl})\Big)\\
\nonumber
&+&L_{ks}\sum_{a\neq j}\frac{1}{\gQ_{ja}}\Big( \delta_{ia}\delta_{al}\delta_{js}\gQ_j + \delta_{ij}\delta_{jl}\delta_{as}\gQ_a\Big)\, .
\eea 
Here the summation in the first line can be extended to include the term with $a=j$ because the latter vanishes. The second line can be conveniently written as 
a matrix element of some $r$-matrix. Namely,  
\bea
\nonumber
\mathbb{B}_{ij,kl}&=&L_{ks}\Big(\rdr_{-ij,sl} -\delta_{ij}\sum_a\rdr_{-ja,sl}\Big)  - \Big(\rdr_{-ij,ks} - \delta_{ij}\sum_a\rdr_{-ja,ks}\Big) L_{sl} \\
\nonumber
&+&L_{ks}\sum_{a\neq b}\frac{\gQ_b}{\gQ_{ab}}(E_{aa}-E_{ab})_{ij}\otimes (E_{ba})_{sl}\, .
\eea 
 In matrix form 
 \bea
 \la{Final_B12}
 \mathbb{B}_{12}=L_2(\rdr_{-12}-a_{12})-(\rdr_{-12}-a_{12})L_2+L_2d_{12}\, ,
 \eea
 where $a_{12}$ is the same matrix as in (\ref{TT_final}) and we introduced 
 \bea
 \la{d12}
d_{12}=\sum_{a\neq b}\frac{\gQ_b}{\gQ_{ab}}(E_{aa}-E_{ab})\otimes E_{ba}\, .
 \eea
We also need 
 \bea
 \nonumber
 \mathbb{B}_{21}=L_1(\rdr_{-21}-a_{21})-(\rdr_{-21}-a_{21})L_1+L_1d_{21}\, ,
 \eea
 Since $\rdr_{-21}=-\rdr_{+12}$, we have 
 \bea
 \la{Final_B21}
 \mathbb{B}_{21}=-L_1(\rdr_{+12}+a_{21})+(\rdr_{+12}+a_{21})L_1+L_1d_{21}\, .
 \eea
Now everything is ready to obtain the bracket (\ref{LL_ap1}). Substituting in (\ref{LL_ap1}) expressions (\ref{TT_final}), (\ref{Final_B12}) and (\ref{Final_B21}), we conclude that (\ref{LL_ap1})
has the structure
\bea
\la{LL_kkss}
\{L_1,L_2\}&=&k_{12}^+L_1L_2+L_1L_2k_{12}^-
+ L_1s_{12}^-L_2+L_2s_{12}^+L_1\, ,
\eea
where the coefficients are 
\bea
\la{coeff_roughf}
\begin{aligned}
k_{12}^+&=r_{\gQ\, 12}+C_{12}+(a_{21}+b_{12}-c_{12})\, ,\\
k_{12}^-&=r_{\gQ\, 12} +d_{12}-d_{21}+(a_{21}+b_{12}-c_{12})\, , \\
s_{12}^+&=-r_{\gQ\, 12}-d_{12}-(a_{21}+b_{12}-c_{12})\, , \\
s_{12}^-&=-r_{\gQ\, 12}-C_{12}+d_{21}-(a_{21}+b_{12}-c_{12})\, .
\end{aligned}
\eea
First, we note that these coefficients satisfy
\bea
\la{sumk}
k^++k^-+s^++s^-=0\, ,
\eea
which guarantees that spectral invariants of $L$ are in involution on the Heisenberg double.
Second, in (\ref{coeff_roughf}) the apparent dependence on the variable $T$ occurs in the single combination $a_{21}+b_{12}-c_{12}$.
To make further progress, consider 
$$
a_{21}=C_{12}a_{12}C_{12}\, ,
$$ 
as $C_{12}$ acts as the permutation. We have, written in components,
\bea
\nonumber
(a_{21})_{ij,kl}&=&C_{im,kn}(a_{12})_{mr,ns}C_{rj,sl}=\delta_{mk}\delta_{in}\Big(\delta_{mr} \sum_a \rdr_{-ma,ns}\Big)\delta_{js}\delta_{r l}\\
\nonumber
&=&\delta_{kl}\sum_a \rdr_{-ka,ij}=-\delta_{kl}\sum_a \rdr_{+ij,ka}\, .
\eea 
Therefore,
\bea
\nonumber
(a_{21}+b_{12})_{ij,kl}&=&-\delta_{kl}\sum_a \rdr_{+ij,ka}+\delta_{kl}\sum_a \rdr_{- ij, k a}=-\delta_{kl}\sum_a C_{ij,ka}\\
\nonumber
&=&-\sum_a\delta_{kl}\delta_{jk}\delta_{ia}=-\sum_{ab} (E_{ab})_{ij}\otimes (E_{bb})_{kl}\, .
\eea
The dependence on $T$ disappears and we find a simple answer 
\bea
\la{ab}
a_{21}+b_{12}=-\sum_{ab} E_{ab}\otimes E_{bb}\, .
\eea
The only $T$-dependence is in the coefficient $c_{12}$. This coefficient cannot be simplified or cancelled, so we leave it in the present form.  
Substituting in (\ref{coeff_roughf}) the matrices (\ref{rgQ}), (\ref{d12}) and (\ref{ab}) and, performing necessary simplifications, we obtain our final 
result for the coefficients of the bracket (\ref{LL_kkss}) 
\bea
\la{final_kkss}
\begin{aligned}
k_{12}^+&=\sum_{a\neq b}\Big(\frac{\gQ_b}{\gQ_{ab}}E_{aa}-\frac{\gQ_a}{\gQ_{ab}}E_{ab}\Big)\otimes (E_{bb}-E_{ba})-c_{12}\, ,\\
k_{12}^-&=\sum_{a\neq b} \frac{\gQ_a}{\gQ_{ab}}  E_{aa}\otimes E_{bb}
-\sum_{a\neq b}\frac{\gQ_a}{\gQ_{ab}}E_{ab}\otimes E_{ba}-\mI\otimes \mI-c_{12} \, ,  \\
s_{12}^+&=-\sum_{a\neq b}\frac{\gQ_a}{\gQ_{ab}} (E_{aa}-E_{ab})\otimes E_{bb}+\mI\otimes \mI+c_{12} \, , \\
s_{12}^-&= -\sum_{a\neq b}\frac{\gQ_b}{\gQ_{ab}} E_{aa}\otimes (E_{bb}-E_{ba})+c_{12}\, .
\end{aligned}
\eea
In fact, the identity matrix $\mI\otimes \mI$ appearing in $k^-$ and $s^+$ can be omitted as it cancels out in the expression (\ref{LL_kkss}).
As was already mentioned, the only $T$-dependence left over is in the term $c_{12}$, namely,  
 \bea
 \la{c12}
 (c_{12})_{ij,kl}=\delta_{ij}\delta_{kl} \sum_{ab}\rdr_{- ia, kb}=\delta_{ij}\delta_{kl} T^{-1}_{im}T^{-1}_{kn}\sum_{ab} r_{- ma, nb} \, .
 \eea
 It is this term which violates the invariance of the bracket (\ref{LL_kkss}) under transformations from the Frobenius group. 
 
To complete our discussion, we consider
  \bea
  \nonumber
 (c_{21})_{ij,kl}=\delta_{ij}\delta_{kl} \sum_{ab}\rdr_{- ka, ib}=-\delta_{ij}\delta_{kl}  \sum_{ab}\rdr_{+ ia,kb}\, .
 \eea
 This gives 
 \bea
 \nonumber
 (c_{21}+c_{12})_{ij,kl}=-\delta_{ij}\delta_{kl} \sum_{ab}C_{ia,kb}=-\delta_{ij}\delta_{kl} \sum_{ab}\delta_{ib}\delta_{ka}=-\delta_{ij}\delta_{kl}=-(\mI\otimes \mI)_{ij,kl}\, ,
 \eea
 or in other words,
 \bea
 \la{prop_c}
 c_{21}+c_{12}=-\mI\otimes \mI\, .
 \eea
 Equation (\ref{prop_c}) leads to the following relations between the coefficiencients  
 \bea
 \la{inv_coeff}
 k_{12}^{+} + k_{21}^{+}=C_{12}-2\, (\mI\otimes \mI)\, , ~~~~~k_{12}^{-} + k_{21}^{-}=-C_{12}\,  , ~~~s_{12}^-=-s^+_{21}\, .
 \eea
Notice that the fact that the right-hand side of the first two expressions is an invariant tensor.  Relations (\ref{inv_coeff})  guarantee that the bracket (\ref{LL_kkss}) is skew-symmetric.

Following similar steps, we can derive the Poisson brackets involving other Frobenius invariants on the Heisenberg double, namely \(W_{ij}\) and \(P_i\) coordinates. Introducing the notations
\begin{align*}
\rr_{\pm}^{hg}&= h_1^{-1}g_2^{-1}\rr_{\pm}h_1 g_2\,,\qquad
(c_{12}^{hg})_{ijkl}=\delta_{ij}\delta_{kl}\sum_{\alpha,\beta} (\rr^{hg}_{-})_{i\alpha k \beta}\, ,
\end{align*} 
which, for Frobenius elements \(g,h\) satisfies
\begin{align*}
c_{21}^{hg}+c_{12}^{gh}\,=\,-\mathbbm{1}\otimes \mathbbm{1}\, ,
\end{align*}
we can write \bea
\label{WPHD}
\begin{aligned}
\{W_1,W_2\}\,=&\,[r_{12},W_1 W_2]+ W_1 \,c_{12}^{UT} \,W_2+  W_2 \, c_{12}^{TU}\,W_1 - W_1 W_2 \,c_{12}^{UU} - c_{12}^{TT} \, W_1 W_2 \\
\{W_1,P_2\}\,=&\,P_2 [\bar r_{12},W_1]+ P_2 W_1(c_{12}^{UT} -  \, c_{12}^{UU}) + P_2 (c_{12}^{TU} -  \, c_{12}^{TT})W_1\\
\{P_1,P_2\}\,=&\,P_1 P_2 \,(c_{12}^{UT}+c_{12}^{TU}-c_{12}^{TT}-c_{12}^{UU})\, ,
\end{aligned}
\eea
where matrices \(r_{12}\) and \(\bar r_{12}\) are defined in \eqref{r_trig}.
The \(c^{hg}\)-like terms in the brackets \eqref{WPHD} are not Frobenius invariants, despite the arguments of the brackets are so, as it happens for \eqref{LL_kkss}. These terms disappear after imposing Dirac constraints in the reduced phase space, as it will explicitly shown for the \(LL\)-bracket in \ref{A:DB}.
\subsection{Dirac bracket} 
\la{A:DB}
Here we outline the construction of the Dirac bracket between the entries of the Lax matrix (\ref{L_on_HD}).
We argue that the contribution to the Dirac bracket coming from the second class constraints has the same matrix structure as (\ref{LL_kkss}) and that this contribution precisely cancels all the 
terms $c_{12}$ in (\ref{final_kkss}), so that the resulting coefficients describing the Dirac bracket on the constraint surface are given by expressions (\ref{r_trig}) in the main text.

We start with the Poisson algebra of the non-abelian moment map
\bea
\la{M_double_1}
\{\MM_1,\MM_2\}=-\rr_+ \MM_1\MM_2- \MM_1\MM_2 \rr_- + \MM_1 \rr_-\MM_2 + \MM_2 \rr_+\MM_1\, .
\eea
This is the Semenov-Tian-Shansky type bracket; it has $\nn$ Casimir functions ${\rm Tr}(\MM^k)$ with $k=1,\ldots, \nn$.
%where 
%$$
%\uprho_{12}=-r_+ \MM_1\MM_2- \MM_1\MM_2r_- + \MM_1 r_-\MM_2 + \MM_2 r_+\MM_1\, .
%$$
%$$
%\uprho_{12}=-M_1^{-1}M_2^{-1}r_+ M_1M_2- r_- + M_2^{-1}r_-M_2 + M_1^{-1}r_+M_1\, .
%$$
%takes values in $\J\otimes \J$. 
%Substituting here the value of $M$ chosen for reduction, see (\ref{M_trigRS})  
On the constraint surface $\mathcal{S}$ the moment map is fixed to the following value
\bea
\la{M_on_const}
\MM=\omega \mI +\beta \e\otimes \e^\t \, , 
%~~~M^{-1}=\frac{1}{\a}\mI-\frac{\beta}{\a^2+n \a\beta}  \e\otimes \e^t \, ,
\eea
see (\ref{M_trigRS}). Substituting this expression into the right-hand side of (\ref{M_double_1}) yields the following answer 
\bea
\la{M_on_S}
\MM_{ij,kl}\equiv \{\MM_{ij},\MM_{kl}\}\Big|_{\mathcal{S}}&=&\beta\Big[\big(\omega^{1-\nn}-\beta(i-\sfrac{1}{2})\big)\delta_{il}-\sfrac{\beta}{2}\delta_{jl}+\beta\,  \Theta(l-j)\\
\nonumber
&&~~~-\big(\omega^{1-\nn}-\beta(j-\sfrac{1}{2})\big)\delta_{jk}+\sfrac{\beta}{2}\delta_{ik}-\beta\,  \Theta(k-i) \Big]\, ,
\eea
where $\Theta$ is the Heaviside step function 
\bea
\Theta(j)=\left\{ \begin{array}{l} 1,~j\geq 0\, , \\ 0,~ j<0\end{array}\right. .
\eea

For any $X\in {\rm Mat}(N,{\mathbb C})$ 
introduce the following quantities
\bea
\la{coeff_t}
\begin{aligned}
t^{(0)}(X)_{ij}&=X_{ij}-\frac{1}{\nn} \sum_{a}X_{aj}-\frac{1}{\nn}\sum_{a}X_{ia}+\frac{1}{\nn^2}\sum_{ab}X_{ab}\, , ~~~~i,j=2,\ldots, \nn,\\
t^{(1)}(X)_j&=\frac{1}{\nn^2}\sum_{ab}X_{ab}-\frac{1}{\nn}\sum_{a}X_{aj}\, , ~~~~~j=2,\ldots, \nn\, ,\\
t^{(2)}(X)_{j}&=\frac{1}{\nn^2} \sum_{ab}X_{ab}-\frac{1}{\nn}\sum_{a}X_{ja}\, , ~~~~j=2,\ldots, \nn\, , \\
t^{(3)}(X) &=\frac{1}{\nn^2}\sum_{ab}X_{ab}\, .
\end{aligned}
\eea 
From these quantities we construct the projectors $\uppi^{(i)}$ that have the following action on $X$
\bea
\la{proj_dec}
\begin{aligned}
&\uppi^{(0)}(X)=\sum_{i,j=2}^{\nn} (E_{11}-E_{i1}-E_{1j}+E_{ij})\, t^{(0)}(X)_{ij}  \, ,~~~
&\uppi^{(1)}(X)=\sum_{j=2}^{\nn} a_j \, t^{(1)}(X)_{j} \, , \\
&\uppi^{(2)}(X)=\sum_{j=2}^{\nn} b_j \, t^{(2)}(X)_{j} \, , ~~~
&\uppi^{(3)}(X)=\sum_{i,j=1}^{\nn}E_{ij}\,  t^{(3)}(X)\, .
\end{aligned}
\eea
where 
\bea
\nonumber
\begin{aligned}
a_j&=\sum_{i=1}^{\nn} (E_{i1}-E_{ij})\, , ~~~~j=2,\ldots, {\nn}\, ,  \\
b_j&=\sum_{i=1}^{\nn} (E_{1i}- E_{ji})\, , ~~~j=2,\ldots, {\nn} \, .
\end{aligned}
\eea
In particular, $\uppi^{(0)}$ projects on the Lie algebra of $\boldsymbol{\digamma}$ and $\uppi^{(3)}$ --
on the one-dimensional dilatation subalgebra ${\mathbb C}^*$.
The completeness condition is 
$$
X=\sum_{k=0}^3 \uppi^{(k)}(X)\, .
$$

From (\ref{M_on_S}) it is readily seen that 
\bea
\nonumber
\{t^{(3)}(\MM),\MM_{kl}\}=\frac{1}{\nn^2}\sum_{ab}\{\MM_{ab},\MM_{kl}\}=0\, .
\eea
Analogously, we find 
\bea
\nonumber
\{t^{(0)}(M)_{ij},\MM_{kl}\}&=\{\MM_{ij}-\frac{1}{\nn} \sum_{a}\MM_{aj}-\frac{1}{\nn}\sum_{a}\MM_{ia}, \MM_{kl}\} =0\, , ~~~i,j=2,\ldots, \nn\, .
\eea
Thus, projections $\uppi^{(0)}(\MM)$ and $\uppi^{(3)}(\MM)$ constitute $(\nn-1)^2+1=\nn^2-2\nn+2$ constraints 
of the first class. Projections $\uppi^{(1)}$ and $\uppi^{(2)}$ yield a non-degenerate matrix of Poisson brackets and, therefore, represent 
$2(\nn-1)$ constraints of the second class. This matrix should be inverted and used to define the corresponding Dirac bracket. 
Even simpler, the matrix (\ref{M_on_S}) has rank $2(\nn-1)$ and we can use any non-degenerate submatrix of this rank to define 
the corresponding Dirac bracket. 

%One way to single out such non-degenerate submatrix is as follows. Assume for definiteness  that $n$ is odd.
%Then from the general bracket (\ref{M_on_S}) we get the following $(n-1)\times (n-1)$ submatrix
%\bea
%G_{ij}\equiv \{\MM_{1i},\MM_{1j}\}=\frac{\beta^2}{2}\left\{ \begin{array}{rc} 1\, , & i<j \, , \\ 0\, , & i=j\, , \\ -1\, , & i>j\, \end{array} \right.\, , ~~~~i,j=2,\ldots , n\, . 
%\eea
%This submatrix is non-degenerate and has the simple inverse 
%\bea
%G^{-1}_{ij}=\frac{2}{\beta^2}\left\{ \begin{array}{rc} (-1)^{i+j}\, , & i<j \, , \\ 0\, , & i=j\, , \\ -(-1)^{i+j}\, , & i>j\, \end{array} \right.\, , ~~~~i,j=2,\ldots , n\, . 
%\eea

Now we derive the Poisson relations between the moment map $\MM$ and the Lax matrix given by (\ref{L_on_HD}). 
First, we compute 
\bea
\{\MM_{ij},T_{kl}\}&=&\frac{\delta T_{kl}}{\delta A_{rs}}\{\MM_{ij},A_{rs}\}=\\
\nonumber
&=&-((\rr_+\MM_1-\MM_1\rr_-)T_2)_{ij,kl}+T_{kl}\sum_a (T_2^{-1}(\rr_+\MM_1-\MM_1\rr_-))_{ij,la}\, .
\eea
Deriving this formula, we have used (\ref{MAB}), as well as the fact that $T\in F$.
Next, we obtain
\bea
\{\MM_{ij},L_{kl}\}&=&L_{kl}\sum_{sp}(T^{-1}_{ls}-T^{-1}_{ks})(\rr_+\MM_1-\MM_1\rr_-)_{ij,sp}\, .
%\nonumber
%&=&L_{kl}\Big((T^{-1}_{lj}-T^{-1}_{kj})\sum_{p}\MM_{ip} +\sum_{sp}(T^{-1}_{ls}-T^{-1}_{ks})[r_+,\MM_1]_{ij,sp}\Big)\, .
\eea
It is clear that the diagonal entries from this expression  of $L$ commute with all the constraints: $\{\MM_{ij},L_{kk}\}=0$,
even without restricting to the constrained surface. 

On the constrained surface where $\MM$ is given by (\ref{M_on_const}), we have 
\bea
\la{ML_br}
\{\MM_{ij},L_{kl}\}\Big|_{\mathcal{S}}&=&
\nonumber
\omega^{1-\nn}L_{kl}(T^{-1}_{lj}-T^{-1}_{kj})+\beta L_{kl}\sum_{sp}(T^{-1}_{ls}-T^{-1}_{ks})\Omega_{ijs}\, .
\eea
Here
\bea
\la{Omega_abc}
\Omega_{ijs}\equiv \sum_p [\rr_+, (\e\otimes \e^t)_1]_{ij,sp}=-\sfrac{1}{2}\delta_{is}-(j-\sfrac{1}{2})\delta_{js}+\Theta(s-i)\, .
\eea
From the explicit expression (\ref{ML_br}) and the fact that $T$ is an element of the Frobenius group, we further deduce that 
\bea
\nonumber
\{t^{(0)}(\MM)_{ij},L_{kl}\}\Big|_{\mathcal{S}}=0 \, , ~~~\{t^{(3)}(\MM)_{ij},L_{kl}\}\Big|_{\mathcal{S}}=0\, .
\eea
In other words, $L$ commutes on the constraint surface with all constraints of the first class, independently on the value of $T$.

With the help of  (\ref{Omega_abc}) we obtain
\bea
\{\MM_{ij},L_{kl}\}\Big|_{\mathcal{S}}&=&
\nonumber
L_{kl}\Big[(\omega^{1-\nn}-\beta(j-\sfrac{1}{2}) )(T^{-1}_{lj}-T^{-1}_{kj})\\
\nonumber
&&\hspace{2cm}+\sfrac{\beta}{2}(T^{-1}_{li}-T^{-1}_{ki})+\beta\sum_{s>i}(T^{-1}_{ls}-T^{-1}_{ks})\Big]\, .
\eea
Taking into account that 
\bea
\nonumber
\sum_{s>i}^{\nn}(T^{-1}_{ls}-T^{-1}_{ks})+\sum_{s<i}^{\nn}(T^{-1}_{ls}-T^{-1}_{ks})+(T^{-1}_{li}-T^{-1}_{ki})=0\, ,
\eea
we can write 
\bea
\{\MM_{ij},L_{kl}\}\Big|_{\mathcal{S}}&=&
\nonumber
L_{kl}\Big[(\omega^{1-\nn}-\beta(j-\sfrac{1}{2}) )(T^{-1}_{lj}-T^{-1}_{kj})\\
\nonumber
&&\hspace{2cm}+\sfrac{\beta}{2}\Big(\sum_{s>i}^{\nn}(T^{-1}_{ls}-T^{-1}_{ks})-\sum_{s<i}^{\nn}(T^{-1}_{ls}-T^{-1}_{ks})\Big) \Big]\, .
\eea
Now we come to the Dirac bracket construction. By picking a non-degenerate submatrix $\Psi$ of the matrix $\MM_{ij,kl}$, we invert it and define the corresponding Dirac bracket 
\bea
\la{LL_D}
\{L_1,L_2\}_{\rm D}=\{L_1,L_2\}-\sum_{I,J=1}^{2\nn-2}\{L_1,\MM_I\} \Psi^{-1}_{IJ}\{\MM_J,L_2 \}\, .
\eea 
Here $I=(ij)$ is a generalised index which we use to label matrix elements of $\MM_{ij,kl}$ that comprise the non-degenerate matrix $\Psi_{IJ}$. To give an example, for 
$\nn=3$ we can take 
as $\Psi$ the following matrix 

\vskip -0.3cm
{\small
\bea
\nonumber
\Psi=\left(\begin{array}{rrrr} \MM_{11,11} & \MM_{11,12} & \MM_{11,13} & \MM_{11,21} \\
                                            \MM_{12,11} & \MM_{12,12} & \MM_{12,13} & \MM_{12,21} \\
                                            \MM_{13,11} & \MM_{12,12}  & \MM_{13,13} & \MM_{13,21} \\
                                            \MM_{21,11} & \MM_{21,12} & \MM_{21,13} & \MM_{21,21} 
                                            \end{array}\right)=\beta
                                  \left(\begin{array}{rrrr} 0 & -\omega-2\b & -\omega-2\b & -\omega-2\b \\
                                            \omega+2\b  & 0 & \sfrac{\beta}{2} & 0\\
                                            \omega+2\b  & -\sfrac{\beta}{2}  & 0 &\omega+\sfrac{3}{2}\b\\
                                            \omega+2\b  & 0 & -\omega-\sfrac{3}{2}\b &0
                                            \end{array}\right) \, .
\eea
}
\normalsize
In particular $\det \Psi=\beta^4(\omega+\beta)^2(\omega+2\beta)^2$. Inverting $\Psi$, we find that 
\bea
\nonumber
\sum_{I,J=1}^{2\nn-2}\{L_1,\MM_I\} \Psi^{-1}_{IJ}\{\MM_J,L_2 \}=k_{{\rm D}12}^+L_1L_2+L_1L_2k_{{\rm D}12}^-
+ L_1s_{{\rm D}12}^-L_2+L_2s_{{\rm D}12}^+L_1\, ,
\eea
that is, the contribution of the second class constraints has precisely the same structure as (\ref{LL_kkss}). Moreover,
the corresponding coefficients are 
\bea
k_{{\rm D}12}^+=-c_{12}\, , ~~~
k_{{\rm D}12}^-=-c_{12}\, , ~~~
s_{{\rm D}12}^+=c_{12}\, , ~~~
s_{{\rm D}12}^-=c_{12}\, , 
\eea
where $c_{12}$ is given by (\ref{c12}). Thus, in (\ref{LL_D}) all the terms $c_{12}$ cancel out.
We have also performed a similar computation for $\nn=4,5,6,7$ with the same result.  An analytic derivation for arbitrary $\nn$
is still missing, although our findings leave little doubt that it holds true.  

In summary, on the reduced phase space the  Dirac bracket between the components of the Lax matrix has the form (\ref{LL_kkss}) with the following coefficients 
\bea
\begin{aligned}
k_{12}^+&=\sum_{a\neq b}\Big(\frac{\gQ_b}{\gQ_{ab}}E_{aa}-\frac{\gQ_a}{\gQ_{ab}}E_{ab}\Big)\otimes (E_{bb}-E_{ba})\, ,\\
k_{12}^-&=\sum_{a\neq b} \frac{\gQ_a}{\gQ_{ab}}  E_{aa}\otimes E_{bb}
-\sum_{a\neq b}\frac{\gQ_a}{\gQ_{ab}}E_{ab}\otimes E_{ba} \, ,  \\
s_{12}^+&=-\sum_{a\neq b}\frac{\gQ_a}{\gQ_{ab}} (E_{aa}-E_{ab})\otimes E_{bb}\, , \\
s_{12}^-&= -\sum_{a\neq b}\frac{\gQ_b}{\gQ_{ab}} E_{aa}\otimes (E_{bb}-E_{ba})\, .
\end{aligned}
\eea
The coefficients have the following properties 
\bea
\la{prop_kkss}
k_{12}^{\pm} + k_{21}^{\pm}=\pm (C_{12}- \mI\otimes\mI)\, , ~~~s_{12}^-=-s^+_{21}\, ,
\eea
which guarantee, in particular, skew-symmetry of (\ref{LL_kkss}).  In addition, they satisfy the relation (\ref{sumk}).
In the main text we present the formula (\ref{LL_kkss}) in the $r$-matrix form (\ref{LL_trig}) with the following identifications 
\bea
\nonumber
k^+=r\, , ~~~s^+=-\bar{r}\, , ~~~k^-=-\mr\, .
\eea

\section{ Derivation of the spectral-dependent $r$-matrices}
\la{app:cauchy}
To determine the $r$-matrices governing the structure (\ref{sdL_t}), we start with
computing the Poisson brackets between the components  of $L(\lambda)$ given by (\ref{spec_dep_L}). 
Applying the Poisson brackets (\ref{bracket_LQ}) and (\ref{LL_trig}), we obtain 
\bea
\nonumber
\{L_1(\lambda),L_2(\mu)\}&=&
 r_{12} L_1 L_2- L_1 L_2\mr_{12} + L_1\bar{r}_{21}L_2- L_2\bar{r}_{12} L_1\, \\
 \nonumber
 &&{\hspace{-3.5cm}}-\frac{1}{\lambda}\Big[ \gQ_1^{-1}r_{12}\gQ_1 L'_1 L_2- L'_1 L_2(\gQ_1^{-1}\mr_{12}\gQ_1-\overC_{12}) + L'_1\gQ_1^{-1}\bar{r}_{21}\gQ_1L_2- L_2(\gQ_1^{-1}\bar{r}_{12}\gQ_1+\overC_{12}) L'_1\Big]\\
 \nonumber
 &&{\hspace{-3.5cm}}-\frac{1}{\mu}\Big[ \gQ_2^{-1}r_{12}\gQ_2 L_1 L'_2- L_1 L'_2(\gQ_2^{-1}\mr_{12}\gQ_2+\overC_{12}) + L_1(\gQ_2^{-1}\bar{r}_{21}\gQ_2+\overC_{12})L'_2- L'_2\gQ_2^{-1}\bar{r}_{12}\gQ_2 L_1\Big]\\
% \nonumber
 %&&-\mu L_1L'_2\overC_{12}+\lambda L'_1L_2\overC_{12}- \lambda L_2\overC_{12}L'_1+\mu L_1\overC_{12}L'_2\\
 \la{A_br1}
 &&{\hspace{-3.5cm}}+\frac{1}{\lambda\mu}\Big[
  \gQ_1^{-1}\gQ_2^{-1}r_{12}\gQ_1\gQ_2 L'_1 L'_2- L'_1 L'_2  \gQ_1^{-1}\gQ_2^{-1}\mr_{12}\gQ_1\gQ_2 
  \\
  \nonumber
  &&
  + L'_1  (\gQ_1^{-1}\gQ_2^{-1}\bar{r}_{21}\gQ_1\gQ_2+\overC_{12})L'_2- L'_2 ( \gQ_1^{-1}\gQ_2^{-1} \bar{r}_{12}\gQ_1\gQ_2+\overC_{12}) L'_1 \Big]\, .
\eea
Further developments are based on the following observation about the properties of the $r$-matrices rotated by $\gQ$'s. 
First, we find that 
\bea
\la{idr1}
\begin{aligned}
&\gQ_1^{-1}r_{12}\gQ_1=r_{12}-\sigma_{12}-C_{12}+\mI\otimes \mI\, , \\
&\gQ_2^{-1}r_{12}\gQ_2=r_{12}+\sigma_{21}+V_{12}-\mI\otimes \mI\, ,\\
&\gQ_1^{-1}\gQ_2^{-1}r_{12}\gQ_1\gQ_2=r_{12}+\sigma_{21}-\sigma_{12}\, ,
\end{aligned}
\eea
where $\sigma_{12}$ is given by (\ref{sigma_mat}) and we introduced 
\bea
\nonumber
V_{12}=\sum_{i,j=1}^{\nn}\frac{\gQ_i}{\gQ_j}E_{ij}\otimes E_{ji}\, .
\eea
Second,
\bea
\la{idr2}
\begin{aligned}
&\gQ_1^{-1}\mr_{12}\gQ_1-\overC_{12}=\mr_{12}-C_{12}\, ,\\
&\gQ_2^{-1}\mr_{12}\gQ_2+\overC_{12}=\mr_{12}+V_{12}\, , \\
&\gQ_1^{-1}\gQ_2^{-1}\mr_{12}\gQ_1\gQ_2=\mr_{12}\, .
\end{aligned}
\eea
Finally,
\bea
\la{idr3}
\begin{aligned}
&\gQ_1^{-1}\bar{r}_{12}\gQ_1+\overC_{12}=\bar{r}_{12}-\sigma_{12}+\mI\otimes \mI\, , \\
&\gQ_2^{-1}\bar{r}_{12}\gQ_2=\bar{r}_{12}\, , \\
&\gQ_1^{-1}\gQ_2^{-1}\bar{r}_{12}\gQ_1\gQ_2+\overC_{12}=\bar{r}_{12}-\sigma_{12}+\mI\otimes \mI\, .
\end{aligned}
\eea
With the help of (\ref{idr1}), (\ref{idr2}) and (\ref{idr3}) the bracket (\ref{A_br1}) turns into 
\bea
\nonumber
\{L_1(\lambda),L_2(\mu)\}&=&
 r_{12} L_1 L_2- L_1 L_2\mr_{12} + L_1\bar{r}_{21}L_2- L_2\bar{r}_{12} L_1\, \\
 \nonumber
 &&{\hspace{-3cm}}-\frac{1}{\lambda}\Big[(r_{12}-\sigma_{12}-C_{12}) L'_1 L_2- L'_1 L_2(\mr_{12}-C_{12}) + L'_1 \bar{r}_{21} L_2- L_2(\bar{r}_{12}-\sigma_{12}) L'_1\Big]\\
 \nonumber
 &&{\hspace{-3cm}}-\frac{1}{\mu}\Big[ (r_{12}+\sigma_{21}) L_1 L'_2- L_1 L'_2 \mr_{12} + L_1(\bar{r}_{21}-\sigma_{21})L'_2- L'_2\bar{r}_{12} L_1\Big]\\
% \nonumber
 %&&-\mu L_1L'_2\overC_{12}+\lambda L'_1L_2\overC_{12}- \lambda L_2\overC_{12}L'_1+\mu L_1\overC_{12}L'_2\\
 \la{LL_spec}
 &&\hspace{-3cm}+\frac{1}{\lambda\mu}\Big[
 (r_{12}-\sigma_{12}+\sigma_{21}) L'_1 L'_2- L'_1 L'_2  \mr_{12}
  \\
 \nonumber
  &&\hspace{2cm}
  + L'_1  (\bar{r}_{21}-\sigma_{21})L'_2- L'_2 ( \bar{r}_{12}-\sigma_{12}) L'_1 \Big]\, .
%  \nonumber
%&& +L'_1\overC_{12}L'_2-L'_2\overC_{12}L'_1
\eea
Notice that the element $V_{12}$ totally decouples from from the right-hand side of (\ref{LL_spec}), as it satisfies an identity
 $$
 V_{12}L_1 L'_2=L_1L'_2V_{12}\, ,
 $$
 which can be straightforwardly verified by computing its
matrix elements, 
\bea
\nonumber
(V_{12}L_1 L'_2)_{mn,kl}=\omega L_{ml}L_{kn}\gQ_l\gQ_k^{-1}=(L_1L'_2V_{12})_{mn,kl}\, .
\eea  
The next progress relies on the identity (\ref{Lax_ident}), {\it i.e.},
\bea
\la{LprL}
L'=L-\frac{1-\omega^N}{N} \e\otimes c^t L\, ,
\eea
and the special (Frobenius) structure of the $r$-matrices. Indeed, from (\ref{LprL}) it follows that 
\bea
\nonumber
(E_{ii}-E_{ij})L'=(E_{ii}-E_{ij})L\, , ~~~\forall\, i,j=1\, ,\ldots, \nn\, .
\eea
This observation immediately shows that 
\bea
\la{reduc1}
\begin{aligned}
&\bar{r}_{12}L'_1=\bar{r}_{12}L_1\, , ~~~~\bar{r}_{21}L'_2=\bar{r}_{21}L_2\, ,\\
&\sigma_{12}L'_1=\sigma_{12}L_1\, ,~~~\sigma_{21}L'_2=\sigma_{21}L_2\, .
\end{aligned}
\eea
Analogously,
\bea
\la{reduc2}
r_{12}L'_2=r_{12}L_2\, , ~~~~r_{21}L'_1=r_{21}L_1\, .
\eea
Owing to the identity (\ref{prop_r}), we then have 
\bea
\nonumber
r_{12}(L'_1-L_1)=(-r_{21}+C_{12}-\mI\otimes \mI)(L'_1-L_1)=(C_{12}-\mI\otimes \mI)(L'_1-L_1)\, ,
\eea
or, in other words,
\bea
\la{reduc3}
r_{12}L'_1=(C_{12}-\mI\otimes\mI)L'_1+(r_{12}-C_{12}+\mI\otimes\mI)L_1\, .
\eea
Thus, to obtain an irreducible expression for the bracket (\ref{LL_spec}), whenever its is possible we will use the reduction 
formulae  (\ref{reduc1}), (\ref{reduc2}) and (\ref{reduc3}) to replace $L'$ with $L$ on the right-hand side of (\ref{LL_spec}).
This replacement leads to the following result
%\bea
%\nonumber
%\{L_1(\lambda),L_2(\mu)\}&=&
% r_{12} L_1 L_2- L_1 L_2\mr_{12} + L_1\bar{r}_{21}L_2- L_2\bar{r}_{12} L_1\, \\
% \nonumber
% &&{\hspace{-3cm}}-\frac{1}{\lambda}\Big[
% (r_{12}-C_{12}+\mI\otimes\mI)L_1L_2+
 %(-\mI\otimes \mI-\sigma_{12}) L'_1 L_2- L'_1 L_2\mr_{12} +C_{12}L_1L'_2+ L'_1 \bar{r}_{21} L_2- L_2(\bar{r}_{12}-\sigma_{12}) L_1
% \Big]\\
 %\nonumber
% &&{\hspace{-3cm}}-\frac{1}{\mu}\Big[ (r_{12}+\sigma_{21}) L_1 L_2- L_1 L'_2 \mr_{12} + L_1(\bar{r}_{21}-\sigma_{21})L_2- L'_2\bar{r}_{12} L_1\Big]\\
% \la{LL_spec}
% &&\hspace{-3cm}+\frac{1}{\lambda\mu}\Big[
% (r_{12}-C_{12}+\mI\otimes \mI) L_1 L_2+(C_{12}+\sigma_{21}-\mI\otimes\mI)L'_1L_2-\sigma_{12}L_1L'_2- L'_1 L'_2  \mr_{12}
%  \\
% \nonumber
%  &&\hspace{2cm}
%  + L'_1  (\bar{r}_{21}-\sigma_{21})L_2- L'_2 ( \bar{r}_{12}-\sigma_{12}) L_1 \Big]\, .
%\eea
\bea
\nonumber
&&\hspace{-0.5cm}\{L_1(\lambda),L_2(\mu)\}=\\
\nonumber
&&~~ =\Big(r_{12}-\frac{1}{\lambda} (r_{12}-\sigma_{12}-C_{12}+\mI\otimes\mI)-\frac{1}{\mu} (r_{12}+\sigma_{21})+\frac{1}{\lambda\mu} (r_{12}-C_{12}+\mI\otimes \mI)\Big)L_1 L_2\\
\nonumber
&&~~ - L_1 L_2\mr_{12} +\frac{1}{\lambda}L'_1 L_2\mr_{12}+\frac{1}{\mu} L_1 L'_2 \mr_{12} -\frac{1}{\lambda\mu}L'_1 L'_2  \mr_{12} \\
\nonumber
&&~~+ L_1\Big(\bar{r}_{21}-\frac{1}{\mu}(\bar{r}_{21}-\sigma_{21})\Big)L_2- L_2\Big(\bar{r}_{12}-\frac{1}{\lambda}(\bar{r}_{12}-\sigma_{12})\Big) L_1\, \\
\nonumber
&&~~
-\frac{1}{\lambda} L'_1\Big( \bar{r}_{21}-\frac{1}{\mu} (\bar{r}_{21}-\sigma_{21}) \Big)L_2
+\frac{1}{\mu} L'_2\Big(\bar{r}_{12} -\frac{1}{\lambda} ( \bar{r}_{12}-\sigma_{12})\Big)L_1 \\
 \la{directL}
&&~~ +
\frac{1}{\lambda}\Big( \mI\otimes \mI +\frac{1}{\mu} (C_{12}+\sigma_{21}-\mI\otimes\mI)\Big)L'_1 L_2  -\frac{1}{\lambda}\Big(C_{12}+\frac{1}{\mu}\sigma_{12}\Big)L_1L'_2 \, .
\eea
We will now search for the spectral dependent $r$-matrices $r^s$ that allow one to present the bracket above in the form 
\bea
\nonumber
\{L_1(\lambda),L_2(\mu)\}&=&
 r^s_{12} L_1(\lambda) L_2(\mu)- L_1(\lambda)  L_2(\mu)\mr_{12}^s \\
\la{sdL}
 &&~~~~~+ L_1(\lambda) \bar{r}_{21}^sL_2(\mu)- L_2(\mu)\bar{r}_{12}^s L_1(\lambda) \, .
\eea
An examination of this expression shows that it involves the following matrices $r_{12}$, $\bar{r}_{12}$, $\bar{r}_{21}$, $\sigma_{12}$, $\sigma_{21}$
and $C_{12}$. There is also the identity matrix $\mI\otimes \mI$ but we ignore its presence for the moment. Thus, the structure of (\ref{directL}) motivates 
to try for the spectral-dependent $r$-matrices the following minimal ansatz
\bea
\nonumber
\begin{aligned}
r^s_{12}&=r_{12}+\alpha \sigma_{12}+\beta \sigma_{21}+\delta C_{12}\, \\
\bar{r}^s_{12}&=\bar{r}_{12}+\delta_{12} \sigma_{12}\, , \\
\bar{r}^s_{21}&=\bar{r}_{21}+\delta_{21} \sigma_{21}\, , \\
\mr^s_{12}&=\mr_{12}+\delta C_{12}\, .
\end{aligned}
\eea
This ansatz depends on five undermined parameters: $\alpha,\beta,\delta,\delta_{12}$ and $\delta_{21}$, which should eventually be expressed via 
$\lambda$ and $\mu$. We then plug this ansatz together with the expression (\ref{spec_dep_L}) for the spectral-dependent Lax matrix into (\ref{sdL}) and,
by using the reduction formulae  (\ref{reduc1}), (\ref{reduc2}) and (\ref{reduc3}), 
 bring the resulting expression to the following irreducible form 
%\bea
%\nonumber
%\begin{aligned}
%&\{L_1(\lambda),L_2(\mu)\}=(r_{12}+\alpha \sigma_{12}+\beta \sigma_{21}) L_1L_2- L_1L_2\mr_{12} + L_1(\bar{r}_{21}+\delta_{21} \sigma_{21})L_2- L_2(\bar{r}_{12}+%\delta_{12} \sigma_{12} ) L_1\\
%&\hspace{-1cm}-\frac{1}{\lambda}\Big[ (r_{12}+\beta \sigma_{21}+\gamma C_{12}) L'_1L_2
%+\alpha \sigma_{12}L_1L_2
%- L'_1L_2\mr_{12} - \gamma C_{12}L_1L'_2 + L'_1(\bar{r}_{21}+\delta_{21} \sigma_{21})L_2- L_2(\bar{r}_{12}+\delta_{12} \sigma_{12}) L_1 \Big]\\
%&\hspace{-1cm}-\frac{1}{\mu}\Big[(\alpha \sigma_{12}+\gamma C_{12})L_1L'_2
%+(r_{12}+\beta \sigma_{21})L_1L_2
%- L_1L'_2\mr_{12}-\gamma C_{12}L'_1L_2  + L_1(\bar{r}_{21}+\delta_{21} \sigma_{21})L_2- L'_2(\bar{r}_{12}+\delta_{12} \sigma_{12})  L_1\Big]\\
%&\hspace{-1cm}+\frac{1}{\lambda\mu}
%\Big[(r_{12}+\beta \sigma_{21}) L'_1L_2+\alpha \sigma_{12}L_1L'_2- L'_1L'_2\mr_{12} + L'_1(\bar{r}_{21}+\delta_{21} \sigma_{21})L_2- L'_2(\bar{r}_{12}+\delta_{12} %\sigma_{12})  L_1\Big]\, .
%\end{aligned}
%\eea
\bea
\nonumber
&&\{L_1(\lambda),L_2(\mu)\}=\\
\nonumber
&&~=\Big[r_{12}+\alpha \sigma_{12}+\beta \sigma_{21}-\frac{1}{\lambda}(r_{12}-C_{12}+\mI\otimes \mI+\alpha\sigma_{12})\\
\nonumber
&&\hspace{4cm}
-\frac{1}{\mu}(r_{12}+\beta \sigma_{21})+\frac{1}{\lambda\mu}(r_{12}-C_{12}+\mI\otimes \mI)\Big] L_1L_2 \\
\nonumber
&&- L_1L_2\mr_{12} +\frac{1}{\lambda}L'_1L_2\mr_{12}+\frac{1}{\mu}L_1L'_2\mr_{12}-\frac{1}{\lambda\mu}L'_1L'_2\mr_{12}\\
\nonumber
&&+ L_1\Big[ (\bar{r}_{21}+\delta_{21} \sigma_{21})-\frac{1}{\mu}(\bar{r}_{21}+\delta_{21} \sigma_{21})\Big]L_2- L_2\Big[(\bar{r}_{12}+\delta_{12} \sigma_{12} )-\frac{1}{\lambda}(\bar{r}_{12}+\delta_{12} \sigma_{12})\Big] L_1\\
\nonumber
&& -\frac{1}{\lambda} L'_1\Big[(\bar{r}_{21}+\delta_{21} \sigma_{21})-\frac{1}{\mu}(\bar{r}_{21}+\delta_{21} \sigma_{21})\Big]L_2
 +\frac{1}{\mu} L'_2\Big[(\bar{r}_{12}+\delta_{12} \sigma_{12})-\frac{1}{\lambda}(\bar{r}_{12}+\delta_{12} \sigma_{12})   \Big]L_1
\\
\nonumber
&& +\Big[-\frac{1}{\lambda} (C_{12}-\mI\otimes \mI+\beta \sigma_{21}+\delta C_{12}) +\frac{1}{\mu}\delta C_{12}+\frac{1}{\lambda\mu}(C_{12}+\beta \sigma_{21}-\mI\otimes \mI) \Big]L'_1L_2
\\
\la{ansatzL}
&&
+\Big[\frac{1}{\lambda} \delta C_{12}-\frac{1}{\mu}(\alpha \sigma_{12}+\delta C_{12})+\frac{1}{\lambda\mu}\alpha \sigma_{12}\Big]L_1L'_2\, .
\eea
Comparison of the first lines of (\ref{directL}) and (\ref{ansatzL}) yields a unique solution for $\alpha$ and $\beta$,
\bea
\nonumber
\alpha=\frac{1}{\lambda-1}\, , ~~~~~\beta=-\frac{1}{\mu-1}\, .
\eea
Comparison of third lines yields 
\bea
\nonumber
\delta_{12}=\frac{1}{\lambda-1}\, , ~~~~~\delta_{21}=\frac{1}{\mu-1}\, ,
\eea
which automatically makes the fourth lines of (\ref{directL}) and (\ref{ansatzL}) equal. Finally, with $\alpha$ and $\beta$ already determined,
comparison of the terms in front of $L'_1L_2$ or $L_1L'_2$ gives an unambiguous solution for $\delta$,
\bea
\nonumber
\delta=\frac{\mu}{\lambda-\mu}\, .
\eea
Thus, we end up with the following expressions for the spectral-dependent $r$-matrices realising the Poisson algebra (\ref{sdL})
\bea
\la{spec_dep_r}
\begin{aligned}
&r_{12}(\lambda,\mu)=\frac{\lambda r_{12}+\mu r_{21}}{\lambda-\mu}+\frac{\sigma_{12}}{\lambda-1} -\frac{\sigma_{21}}{\mu-1} +\frac{\mu}{\lambda-\mu}\mI\otimes \mI\, ,\\
&\bar{r}_{12}(\lambda)=\bar{r}_{12}+\frac{\sigma_{12}}{\lambda-1} \, , \\
&\mr_{12}(\lambda,\mu)=r_{12}(\lambda,\mu)+\bar{r}_{21}(\mu)-\bar{r}_{12}(\lambda)=\frac{\lambda \mr_{12}+\mu \mr_{21}}{\lambda-\mu}+\frac{\mu}{\lambda-\mu}\mI\otimes \mI\, ,
\end{aligned}
\eea
where we used the relation (\ref{prop_r}) to bring the result to a more symmetric form. Finally, using the shift symmetry 
(\ref{shift_sym}),  we can omit in (\ref{spec_dep_r}) the terms proportional to the 
identity matrix, obtaining a slightly simpler solution (\ref{spec_dep_r_final}).

% \begin{figure}[htb]
% \begin{center}
%   \includegraphics[scale=0.2]{....pdf}
% \end{center}
% \caption{Captions    }
% \label{Label}
% \end{figure}


\begin{thebibliography}{0}

%\cite{Ruijsenaars:1986vq}
\bibitem{Ruijsenaars:1986vq}
  S.~N.~M.~Ruijsenaars and H.~Schneider,
  ``A New Class of Integrable Systems and Its Relation to Solitons,''
  Annals Phys.\  {\bf 170} (1986) 370.
  doi:10.1016/0003-4916(86)90097-7
  %%CITATION = doi:10.1016/0003-4916(86)90097-7;%%
  %182 citations counted in INSPIRE as of 13 Feb 2019
  
%\cite{Ruijsenaars:1986pp}
\bibitem{Ruijsenaars:1986pp}
  S.~N.~M.~Ruijsenaars,
  ``Complete Integrability of Relativistic Calogero-moser Systems and Elliptic Function Identities,''
  Commun.\ Math.\ Phys.\  {\bf 110} (1987) 191.
  doi:10.1007/BF01207363
  %%CITATION = doi:10.1007/BF01207363;%%
  %168 citations counted in INSPIRE as of 13 Feb 2019  

%\cite{Feher:2008moa}
\bibitem{Feher:2008moa}
  L.~Feh\'{e}r and C.~Klimcik,
  %``Poisson-Lie generalization of the Kazhdan-Kostant-Sternberg reduction,''
  Lett.\ Math.\ Phys.\  {\bf 87} (2009) 125
  doi:10.1007/s11005-009-0298-3
  [arXiv:0809.1509 [math-ph]].
  %%CITATION = doi:10.1007/s11005-009-0298-3;%%
  %8 citations counted in INSPIRE as of 22 Feb 2019
  %\cite{Feher:2009kp}
\bibitem{Feher:2009kp}
  L.~Feh\'{e}r and C.~Klimcik,
  %``Poisson-Lie interpretation of trigonometric Ruijsenaars duality,''
  Commun.\ Math.\ Phys.\  {\bf 301} (2011) 55
  doi:10.1007/s00220-010-1140-6
  [arXiv:0906.4198 [math-ph]].
  %%CITATION = doi:10.1007/s00220-010-1140-6;%%
  %25 citations counted in INSPIRE as of 22 Feb 2019
  
%\cite{Feher:2016nta}
\bibitem{Feher:2016nta}
  L.~Feh\'{e}r and T.~F.~G\"{o}rbe,
  ``The full phase space of a model in the Calogero-Ruijsenaars family,''
  J.\ Geom.\ Phys.\  {\bf 115} (2017) 139
  doi:10.1016/j.geomphys.2016.04.018
  [arXiv:1603.02877 [math-ph]].
  %%CITATION = doi:10.1016/j.geomphys.2016.04.018;%%
  %5 citations counted in INSPIRE as of 13 Feb 2019
  
  %\cite{Feher:2018pmu}
\bibitem{Feher:2018pmu}
  L.~Feh\'{e}r,
  ``Poisson-Lie analogues of spin Sutherland models,''
  arXiv:1809.01529 [math-ph].
  %%CITATION = ARXIV:1809.01529;%%
  %3 citations counted in INSPIRE as of 13 Feb 2019


\bibitem[Chalykh(2018)]{2018arXiv180401766C} 
   O.~Chalykh, 
  ``Quantum Lax pairs via Dunkl and Cherednik operators", 2018,
  arXiv:1804.01766.


%\cite{Chalykh:2018wce}
\bibitem{Chalykh:2018wce}
  O.~Chalykh and M.~Fairon,
  ``On the Hamiltonian formulation of the trigonometric spin Ruijsenaars-Schneider system,''
  arXiv:1811.08727 [math-ph].
  %%CITATION = ARXIV:1811.08727;%%
  %1 citations counted in INSPIRE as of 13 Feb 2019

%\cite{Zabrodin:2014bpa}
\bibitem{Zabrodin:2014bpa}
  A.~Zabrodin,
  %``Quantum spin chains and integrable many-body systems of classical mechanics,''
  arXiv:1409.4099 [math-ph].
  %%CITATION = ARXIV:1409.4099;%%
  %3 citations counted in INSPIRE as of 15 Feb 2019
  
%\cite{Grekov:2018htz}
\bibitem{Grekov:2018htz} 
  A.~Grekov, A.~Zabrodin and A.~Zotov,
  %``Supersymmetric extension of qKZ-Ruijsenaars correspondence,''
  Nucl.\ Phys.\ B {\bf 939}, 174 (2019)
  doi:10.1016/j.nuclphysb.2018.12.014
  [arXiv:1810.12658 [math-ph]].
  %%CITATION = doi:10.1016/j.nuclphysb.2018.12.014;%%
%\cite{Isachenkov:2017qgn}
\bibitem{Isachenkov:2017qgn}
  M.~Isachenkov and V.~Schomerus,
  %``Integrability of conformal blocks. Part I. Calogero-Sutherland scattering theory,''
  JHEP {\bf 1807} (2018) 180
  doi:10.1007/JHEP07(2018)180
  [arXiv:1711.06609 [hep-th]].
  %%CITATION = doi:10.1007/JHEP07(2018)180;%%
  %11 citations counted in INSPIRE as of 14 Feb 2019

%\cite{Babelon:1990qk}
\bibitem{Babelon:1990qk}
  O.~Babelon and C.~M.~Viallet,
  ``Hamiltonian Structures and Lax Equations,''
  Phys.\ Lett.\ B {\bf 237} (1990) 411.
  doi:10.1016/0370-2693(90)91198-K
  %%CITATION = doi:10.1016/0370-2693(90)91198-K;%%
  %87 citations counted in INSPIRE as of 13 Feb 2019

\bibitem{Mac}
I.~G.~Macdonald, {\it Symmetric functions and Hall polynomials}. 
Oxford: Clarendon Press; New York: Oxford University Press, 180 p., 1995.
%%CITATION = INSPIRE-383627;%%"

\bibitem{Has}
K. Hasegawa, ``Ruijsenaars' Commuting Difference Operators as Commuting Transfer Matrices",
 Comm. Math. Phys. (1997) 187: 289. https://doi.org/10.1007/s002200050137
 
 
%\cite{Antonov:1997zc}
\bibitem{Antonov:1997zc}
  A.~Antonov, K.~Hasegawa and A.~Zabrodin,
  ``On trigonometric intertwining vectors and nondynamical R matrix for the Ruijsenaars model,''
  Nucl.\ Phys.\ B {\bf 503} (1997) 747
  doi:10.1016/S0550-3213(97)00520-8
  [hep-th/9704074].
  %%CITATION = doi:10.1016/S0550-3213(97)00520-8;%%
  %22 citations counted in INSPIRE as of 13 Feb 2019


 \bibitem{KKS}
 D. Kazhdan, B. Kostant, and S. Sternberg, ``Hamiltonian group actions and dynamical
systems of calogero type,"  Communications on Pure and Applied Mathematics,
31 (4) 481-507, 1978.
 
 %\cite{Gorsky:1993dq}
\bibitem{Gorsky:1993dq}
  A.~Gorsky and N.~Nekrasov,
  ``Relativistic Calogero-Moser model as gauged WZW theory,''
  Nucl.\ Phys.\ B {\bf 436} (1995) 582
  doi:10.1016/0550-3213(94)00499-5
  [hep-th/9401017].
  %%CITATION = doi:10.1016/0550-3213(94)00499-5;%%
  %87 citations counted in INSPIRE as of 13 Feb 2019
 
 %\cite{Gorsky:1994dj}
\bibitem{Gorsky:1994dj}
  A.~Gorsky and N.~Nekrasov,
  ``Elliptic Calogero-Moser system from two-dimensional current algebra,''
  hep-th/9401021.
  %%CITATION = HEP-TH/9401021;%%
  %93 citations counted in INSPIRE as of 13 Feb 2019
  
  %\cite{Arutyunov:1996cmb}
\bibitem{Arutyunov:1996cmb}
  G.~E.~Arutyunov and S.~A.~Frolov,
  ``Quantum Dynamical R-matrices and Quantum Frobenius Group,''
  Commun.\ Math.\ Phys.\  {\bf 191} (1998) 15
  doi:10.1007/s002200050259
  [q-alg/9610009].
  %%CITATION = doi:10.1007/s002200050259;%%
  %10 citations counted in INSPIRE as of 11 Feb 2019
%\cite{Suris:1996mg}


%\cite{Arutyunov:1996vy}
\bibitem{Arutyunov:1996vy}
  G.~E.~Arutyunov, S.~A.~Frolov and P.~B.~Medvedev,
  ``Elliptic Ruijsenaars-Schneider model from the cotangent bundle over the two-dimensional current group,''
  J.\ Math.\ Phys.\  {\bf 38} (1997) 5682
  doi:10.1063/1.532160
  [hep-th/9608013].
  %%CITATION = doi:10.1063/1.532160;%%
  %20 citations counted in INSPIRE as of 11 Feb 2019
  
  %\cite{Arutyunov:1997ey}
\bibitem{Arutyunov:1997ey}
  G.~E.~Arutyunov and S.~A.~Frolov,
  ``On Hamiltonian structure of the spin Ruijsenaars-Schneider model,''
  J.\ Phys.\ A {\bf 31} (1998) 4203
  doi:10.1088/0305-4470/31/18/010
  [hep-th/9703119].
  %%CITATION = doi:10.1088/0305-4470/31/18/010;%%
  %7 citations counted in INSPIRE as of 14 Feb 2019


%\cite{Arutyunov:1996uw}
\bibitem{Arutyunov:1996uw}
  G.~E.~Arutyunov, S.~A.~Frolov and P.~B.~Medvedev,
  ``Elliptic Ruijsenaars-Schneider model via the Poisson reduction of the affine Heisenberg double,''
  J.\ Phys.\ A {\bf 30} (1997) 5051
  doi:10.1088/0305-4470/30/14/016
  [hep-th/9607170].
  %%CITATION = doi:10.1088/0305-4470/30/14/016;%%
  %14 citations counted in INSPIRE as of 11 Feb 2019

\bibitem{ACF} G.~E.~Arutyunov, L.~Chekhov and S.~Frolov,
 ``$R$-Matrix Quantization of the Elliptic Ruijsenaars-Schneider Model,"
Commun. Math. Phys., {\bf 192} (1998) 405--432.  


%\cite{SemenovTianShansky:1985my}
\bibitem{SemenovTianShansky:1985my}
  M.~A.~Semenov-Tian-Shansky,
  ``Dressing transformations and Poisson group actions,''
  Publ.\ Res.\ Inst.\ Math.\ Sci.\ Kyoto {\bf 21} (1985) 1237.
  doi:10.2977/prims/1195178514
  %%CITATION = doi:10.2977/prims/1195178514;%%
  %142 citations counted in INSPIRE as of 11 Feb 2019
\bibitem{Suris:1996mg}
  Y.~B.~Suris,
  %``Why are the Ruijsenaars-Schneider and the Calogero-Moser hierarchies governed by the same tau matrix?,''
  Phys.\ Lett.\ A {\bf 225} (1997) 253
  doi:10.1016/S0375-9601(96)00897-3
  [hep-th/9602160].
  %%CITATION = doi:10.1016/S0375-9601(96)00897-3;%%
  %19 citations counted in INSPIRE as of 22 Feb 2019
  
  %\cite{Avan:1995dk}
\bibitem{Avan:1995dk}
  J.~Avan and G.~Rollet,
  %``The Classical r matrix for the relativistic Ruijsenaars-Schneider system,''
  Phys.\ Lett.\ A {\bf 212} (1996) 50
  doi:10.1016/0375-9601(96)00068-0
  [hep-th/9510166].
  %%CITATION = doi:10.1016/0375-9601(96)00068-0;%%
  %8 citations counted in INSPIRE as of 22 Feb 2019
  
  %\cite{Babelon:1993bx}
\bibitem{Babelon:1993bx}
  O.~Babelon and D.~Bernard,
  %``The Sine-Gordon solitons as a N body problem,''
  Phys.\ Lett.\ B {\bf 317} (1993) 363
  doi:10.1016/0370-2693(93)91009-C
  [hep-th/9309154].
  %%CITATION = doi:10.1016/0370-2693(93)91009-C;%%
  %60 citations counted in INSPIRE as of 22 Feb 2019
  
 %\cite{Nijhoff:1996pr}
\bibitem{Nijhoff:1996pr}
  F.~W.~Nijhoff, V.~B.~Kuznetsov, E.~K.~Sklyanin and O.~Ragnisco,
  %``Dynamical r matrix for the elliptic Ruijsenaars-Schneider system,''
  J.\ Phys.\ A {\bf 29} (1996) L333
  doi:10.1088/0305-4470/29/13/005
  [solv-int/9603006].
  %%CITATION = doi:10.1088/0305-4470/29/13/005;%%
  %15 citations counted in INSPIRE as of 22 Feb 2019
  
  %\cite{Sklyanin:1993uj}
\bibitem{Sklyanin:1993uj}
  E.~K.~Sklyanin,
  %``Dynamical r matrices for the elliptic Calogero-Moser model,''
  Alg.\ Anal.\  {\bf 6} (1994) no.2,  227
   [St.\ Petersburg Math.\ J.\  {\bf 6} (1995) 397]
  [hep-th/9308060].
  %%CITATION = HEP-TH/9308060;%%
  %46 citations counted in INSPIRE as of 22 Feb 2019

%\cite{Fock:1998nu}
\bibitem{Fock:1998nu}
  V.~V.~Fock and A.~A.~Rosly,
  ``Poisson structure on moduli of flat connections on Riemann surfaces and r matrix,''
  Am.\ Math.\ Soc.\ Transl.\  {\bf 191} (1999) 67
  [math/9802054 [math-qa]].
  %%CITATION = MATH/9802054;%%
  %78 citations counted in INSPIRE as of 14 Feb 2019
  
    
  %\cite{Gervais:1983ry}
\bibitem{Gervais:1983ry}
  J.~L.~Gervais and A.~Neveu,
  ``Novel Triangle Relation and Absence of Tachyons in Liouville String Field Theory,''
  Nucl.\ Phys.\ B {\bf 238} (1984) 125.
  doi:10.1016/0550-3213(84)90469-3
  %%CITATION = doi:10.1016/0550-3213(84)90469-3;%%
  %288 citations counted in INSPIRE as of 11 Feb 2019
  
  %\cite{Felder:1994pb}
\bibitem{Felder:1994pb}
  G.~Felder,
  ``Conformal field theory and integrable systems associated to elliptic curves,''
  hep-th/9407154.
  %%CITATION = HEP-TH/9407154;%%
  %44 citations counted in INSPIRE as of 11 Feb 2019

%\cite{SemenovTianShansky:1993ws}
\bibitem{SemenovTianShansky:1993ws}
  M.~A.~Semenov-Tian-Shansky,
  ``Poisson Lie groups, quantum duality principle, and the quantum double,''
  Theor.\ Math.\ Phys.\  {\bf 93} (1992) 1292
   [Teor.\ Mat.\ Fiz.\  {\bf 93N2} (1992) 302]
  doi:10.1007/BF01083527
  [hep-th/9304042].
  %%CITATION = doi:10.1007/BF01083527;%%
  %36 citations counted in INSPIRE as of 14 Feb 2019

%\cite{Avan:2003ke}
\bibitem{Avan:2003ke} 
  J.~Avan and A.~Doikou,
  %``Commuting quantum traces for reflection algebras,''
  J.\ Phys.\ A {\bf 37}, 1603 (2004)
  doi:10.1088/0305-4470/37/5/010
  [math/0305424 [math-qa]].
  %%CITATION = doi:10.1088/0305-4470/37/5/010;%%
  %3 citations counted in INSPIRE as of 19 Mar 2019
  
%\cite{Nagy:2004jv}
\bibitem{Nagy:2004jv} 
  Z.~Nagy, J.~Avan, A.~Doikou and G.~Rollet,
  %``Commuting quantum traces for quadratic algebras,''
  J.\ Math.\ Phys.\  {\bf 46}, 083516 (2005)
  doi:10.1063/1.2007587
  [math/0403246 [math-qa]].
  %%CITATION = doi:10.1063/1.2007587;%%
  %7 citations counted in INSPIRE as of 19 Mar 2019

\end{thebibliography}
\end{document}